\documentclass[aps,cqg,twocolumn,floatfix,preprintnumbers,altaffilletter,superscriptaddress]{revtex4-1}
\usepackage{graphicx,url,amssymb,amsmath,rotating,color,units,wasysym,epsfig,multirow,epstopdf,subcaption,xcolor}
\usepackage[colorlinks,urlcolor=blue,citecolor=blue,linkcolor=blue]{hyperref}
\usepackage{soul,xcolor}

\newcommand{\appropto}{\mathrel{\vcenter{
  \offinterlineskip\halign{\hfil$##$\cr
    \propto\cr\noalign{\kern2pt}\sim\cr\noalign{\kern-2pt}}}}}

\begin{document}
\title{Gravitational-wave inference in the catalog era: evolving priors and marginal events}

\author{Shanika Galaudage}
\affiliation{Monash Centre for Astrophysics, School of Physics and Astronomy, Monash University, VIC 3800, Australia}
\affiliation{OzGrav: The ARC Centre of Excellence for Gravitational-Wave Discovery, Clayton, VIC 3800, Australia}

\author{Colm Talbot}
\affiliation{Monash Centre for Astrophysics, School of Physics and Astronomy, Monash University, VIC 3800, Australia}
\affiliation{OzGrav: The ARC Centre of Excellence for Gravitational-Wave Discovery, Clayton, VIC 3800, Australia}

\author{Eric Thrane}
\affiliation{Monash Centre for Astrophysics, School of Physics and Astronomy, Monash University, VIC 3800, Australia}
\affiliation{OzGrav: The ARC Centre of Excellence for Gravitational-Wave Discovery, Clayton, VIC 3800, Australia}

\begin{abstract}
As the number of gravitational-wave transient detections grows, the inclusion of marginally significant events in gravitational-wave catalogs will lead to increasing contamination from false positives.
In this paper, we address the question of how to carry out population studies in light of the fact that some fraction of marginally significant gravitational-wave events are of terrestrial origin.
We show that previously published estimates of $p_\text{astro}$, the probability that an event is of astrophysical origin, imply an effective noise likelihood, which can be used to take into account the uncertain origin of marginal events in population studies.
We derive a formalism to carry out population studies with ambiguous gravitational-wave events.
We demonstrate this formalism using events from the LIGO/Virgo Gravitational-Wave Transient Catalog 1 (GWTC-1) as well as events from the Venumadhav et al. ``IAS catalog.''
We derive posterior distributions for population parameters and discuss how they change when we take into account $p_\text{astro}$.
We provide updated individual-event posterior distributions by including population information.
\end{abstract}

\maketitle

\section{Introduction}
Gravitational-wave astronomy is providing us with a new way to probe compact objects. 
Gravitational-wave signals from coalescing binary black holes are typically described by fifteen parameters~\footnote{Eccentric systems require an additional parameter.}: eight intrinsic parameters describing the masses and spins of the black holes and seven extrinsic parameters describing their orientation and location in space and time.
Signals from binary neutron star mergers are described by additional tidal parameters.

Gravitational-wave astronomers infer these parameters using Bayesian inference.
Bayesian parameter estimation software is used to construct probability distributions for each parameter using stochastic samplers such as nested samplers (e.g., \cite{dynesty}) or Markov Chain Monte Carlo (e.g., \cite{ptemcee}), or, alternatively, likelihood interpolation methods (e.g., \cite{Pankow2015, lange2018}).
These distributions allow us to probe the formation mechanism of compact binaries~\cite{Zevin2017, Lower2018, Wysocki2018, RomeroShaw2019,salvo,Stevenson,Talbot2017,GerosaBerti,Farr2017,ratepopO2}, the fate of massive stars~\cite{Taylor2018,Fishbach2017}, and the nature of matter at extreme densities~\cite{EOS}, and cosmological parameters~\cite{Abbott2017_Hubble, Chen2018, Abbott2019_Hubble}, to name a few highlights.

During the first and second observing runs of Advanced LIGO  \citep{AdLIGO} and Virgo \citep{Virgo} (called O1 and O2 respectively), there were eleven gravitational-wave detections \cite{gwtc1}. The LIGO/Virgo catalog (GWTC-1) includes ten binary black hole detections and one from a binary neutron star detection \cite{GW170817}. Independent analyses of LIGO/Virgo open data~\cite{losc} have yielded additional catalogs~\cite{ogc1, ogc2, ias2}, which confirm many of the original detections, while identifying 9 additional events and one candidate event that are more likely than not to be astrophysical in origin.
These catalogs have facilitated population studies of compact binary mergers, which are beginning to shed light on the nature of stellar evolution and the formation mechanisms of compact binaries \citep{Stevenson2015, Mandel2016, Belczynski2017, Miyamoto2017, Mandel2017, Fishbach2017, Farr2017, Barrett2018, Taylor2018, Talbot2017, Talbot2018, Bouffanais2019, ratepopO2}.

The latest LIGO/Virgo observing run (O3) is underway and gravitational-wave detections are being recorded at a rate of roughly one event a week. As gravitational-wave transient catalogs grow, we expect an increasing number of marginal events to be included, ultimately leading to low-level contamination of false-positive events. 
If we fail to take into account the fact that some events in the growing catalogs are likely of terrestrial origin, we are liable to draw faulty conclusions about the population properties of compact binaries.

Gravitational-wave transient candidates are classified by $p_\text{astro}$, the probability that the event is of astrophysical origin~\cite{pastro}.
In~\cite{gwtc1}, a threshold of $p_\text{astro}>0.5$ is applied in order to determine if a candidate is included in GWTC-1.
However, the population analysis of GWTC-1~\cite{ratepopO2} does not take into account the ambiguous nature of some events.
This is problematic considering one of the most massive events is observed with $p_\text{astro}$ ranging from 0.52-0.98, depending on the search pipeline~\cite{gwtc1}.
Subsequent detection claims in~\cite{ias2} include events with unusually large spins, but borderline values of $p_\text{astro}$.

In this paper, we derive a formalism that takes into account the uncertain origin of gravitational-wave detections using $p_\text{astro}$, building on work from~\cite{fgmc} and~\cite{digging}.
We demonstrate this formalism using events in the GWTC-1~\cite{gwtc1} and ``IAS''~\cite{ias2,ias3,ias4} catalogs.
The remainder of this paper is organized as follows.
In Section~\ref{formalism} we derive a formalism to carry out population studies with ambiguous events, culminating in Eq.~\ref{eq:Upsilon}, which provides the likelihood function for $N$ gravitational-wave events, taking into account $p_\text{astro}$, selection effects, and merger rate.
In Section~\ref{demonstration} we apply our formalism to events in the GWTC-1 and IAS catalogs. 
We investigate how the uncertain origin of some events affects the astrophysical interpretation of these catalogs. 
In Section~\ref{reweight}, we present updated posterior distributions for events in GWTC-1 and the IAS catalog using prior distributions informed by the population Section~\ref{demonstration}.
In Section~\ref{discussion}, we provide closing thoughts and discuss possible avenues for future work.

\section{Formalism}\label{formalism}
In this section, we derive a formalism that uses the $p_\text{astro}$ values of each gravitational-wave event to inform our population analyses. 
In~\ref{EffectiveNoise}, we show how $p_\text{astro}$ implies an effective noise evidence.
This effective noise evidence can be used to construct a more general ``astro'' likelihood function, which allows for a candidate event to be either astrophysical or terrestrial in origin.
In~\ref{posteriors}, we apply the astro likelihood to study individual events with ambiguous origin.
In subsection~\ref{populations}, we extend the astro likelihood from single events to ensembles of $N$ events.
In~\ref{selection}, we take into account selection effects.
In~\ref{rates}, we take into account Poisson counting statistics and merger rate.
In~\ref{everything_together}, we combine the results from the previous subsections to derive a final formula (Eq.~\ref{eq:Upsilon}) for population inference with ambiguous events and selection effects.
Finally, in~\ref{update}, we describe how to update initial estimates of $p_\text{astro}$ using the results of population inference.

\subsection{The effective noise evidence}\label{EffectiveNoise}
We begin with likelihood function which includes the possibility of the signal being astrophysical or terrestrial:
\begin{align}
    {\cal L}_\text{astro}(d|\theta) = \xi{\cal L}(d|\theta) + 
    (1-\xi){\cal L}(d|\O) .
\end{align}
Here $\xi$ is the prior for the astrophysical hypothesis and $1-\xi$ is the prior for the terrestrial hypothesis. 
Meanwhile, ${\cal L}{(d|\theta)}$ is the (usual) likelihood of the data $d$ given signal parameter $\theta$, and ${\cal L}(d|\O)$ is the likelihood of the data given noise.

We take the signal likelihood to be Gaussian
\begin{align}\label{eq:signal_likelihood}
    \log{\cal L}(d|\theta) = -\frac{1}{2}\Big[
    \langle d,d \rangle
    -2\langle d,\mu(\theta) \rangle
    +\langle\mu(\theta),\mu(\theta)\rangle
    \Big] ,
\end{align}
where we use the inner product convention
\begin{align}
    \langle a,b \rangle \equiv 4\Delta f \sum_j
    \Re \left(\frac{a*_j b_j}{P_j}\right) .
\end{align}
Here, the sum over $j$ denotes a sum over frequency bins with width $\Delta f$, while $P_j$ is the noise power spectral density.
In many applications of gravitational-wave inference~\cite{lalinference,bilby,gwtc1}, the noise likelihood is taken to be Gaussian as well
\begin{align}
    \log{\cal L}(d|\O) = -\frac{1}{2} \langle d,d \rangle .
\end{align}
This is probably a reasonable approximation for unambiguous detections and is consistent with the signal likelihood in Eq.~\ref{eq:signal_likelihood}.
However, the Gaussian noise assumption is likely to break down for marginal events.
This is clear when we consider the fact that detection significance is determined using time slides (and other bootstrap methods) due to the unreliability of the Gaussian noise assumption~\cite{pycbc,gstlal,spiir}. Thus we must construct some ``effective noise likelihood''.

Recent work describes procedures for calculating significance using $p_\text{astro}$, the probability that a trigger is of astrophysical origin~\cite{pastro,gwtc1,bcr2,BCR1}; for more discussion, see Appendix~\ref{pastro}.
Assuming $p_\text{astro}$ satisfies this property, it can be interpreted as
\begin{align}\label{eq:pastro}
    p_\text{astro} = 
    \frac{\xi{\cal L}_\theta(d)}{\xi{\cal L}_\theta(d) + (1-\xi){\cal L}_\text{eff}(d|\O)} .
\end{align}
Here
\begin{align}
    {\cal L}_\theta(d) = \int d\theta \, {\cal L}(d|\theta)
    \pi(\theta) ,
\end{align}
is the Bayesian evidence for the astrophysical hypothesis.
We rearrange this equation to solve for ``the effective noise evidence''
\begin{align}\label{eq:effective_noise}
    {\cal L}_\text{eff}(d|\O) \equiv 
    \left(\frac{\xi}{1-\xi}\right)
    \left(\frac{1-p_\text{astro}}{p_\text{astro}}\right) 
    {\cal L}_\theta(d) .
\end{align}

It is instructive to study the limiting behavior of the effective noise evidence.
For triggers with $p_\text{astro}\approx50\%$ (and given equal prior support for each hypothesis), the effective noise evidence is equal to the signal evidence---in agreement with intuition.
It is also worth pausing to ask: is it inconsistent to adopt a Gaussian likelihood for the signal while adopting a different model for the noise?
We argue that this is a reasonable model for LIGO/Virgo data.
Both signals and glitches are relatively rare.
If we assume there is a signal in the data, it is a good approximation to assume that there is probably no glitch present, and so the Gaussian noise approximation is suitable.
However, when choosing a suitable noise likelihood, we are interested precisely in the rare glitches that give rise to spurious triggers.
Thus, a non-Gaussian likelihood (inferred using $p_\text{astro}$) is suitable.

While we are somewhat quick to dismiss the chance of a simultaneous signal and glitch, we acknowledge that such a coincidence occurred during the observation of GW170817~\cite{GW170817}.
The approach taken at the time was to excise the glitch from the data~\cite{GW170817_deglitch}, which is tantamount to the construction of a boutique signal likelihood function.
In other words, signals on top of glitches are currently treated on a case-by-case basis.
It is interesting to consider how one might develop a systematic signal+glitch likelihood.
We touch on this again in Section~\ref{discussion}.

Putting everything together, we rewrite the astro likelihood in terms of $p_\text{astro}$:
\begin{align}\label{eq:Lastro}
    {\cal L}_\text{astro}(d|\theta) \equiv & \xi {\cal L}(d|\theta) + 
    (1-\xi){\cal L}_\text{eff}(d|\O) \nonumber\\
    = & \xi \Big( {\cal L}(d|\theta) + 
    \frac{1-p_\text{astro}}{p_\text{astro}} {\cal L}_\theta(d) \Big) .
\end{align}
Our results so far are similar to the findings from~\cite{digging}, which employ a mixture model likelihood function and a specific model for the noise likelihood.
Our approach, however, does not require us to select a noise model.
Rather, in our formulation, the noise model is hidden within $p_\text{astro}$.
An advantage of this more general approach is that one can take values of $p_\text{astro}$ at face value without needing to know the precise recipe for how each value of $p_\text{astro}$ is calculated.
This allows the population inference problem to be framed in a way that is decoupled from the noise model.
It also enables analysis of candidate events identified by different pipelines using different noise models~\cite{ias1}.

\subsection{Posteriors for ambiguous events}\label{posteriors}
Using the astro likelihood from Eq.~\ref{eq:Lastro}, we construct a  posterior for $\theta$
\begin{align}
    P_\text{astro}(\theta|d) = \frac{{\cal L}_\text{astro}(d|\theta) \pi(\theta)}
    {{\cal Z}_\text{astro}} ,
\end{align}
where
\begin{align}\label{eq:marg_likelihood_astro}
    {\cal Z}_\text{astro} = & 
    \int d\theta \, \pi(\theta) 
    {\cal L}_\text{astro}(d|\theta) \\
    = & \frac{\xi}{p_\text{astro}}{\cal L_\theta}(d)
\end{align}
We use capital $P$ for posteriors to avoid confusion with $p_\text{astro}$.

Substituting, we obtain the following expression for the joint posterior
\begin{align}
    P_\text{astro}(\theta|d) = 
    p_\text{astro}\frac{{\cal L}(d|\theta)\pi(\theta)}
    {{\cal L}_{\theta}} + \big(1-p_\text{astro}\big)\pi(\theta).
\end{align}
Of course, the term
\begin{align}
    P(\theta|d) = \frac{{\cal L}(d|\theta)\pi(\theta)}{{\cal L}_{\theta}} ,
\end{align}
is just the usual expression for the posterior of $\theta$ given the astrophysical hypothesis.
Thus the astro likelihood can be rewritten like so
\begin{align}\label{eq:p_astro_theta}
    P_\text{astro}(\theta|d) = p_\text{astro}\, P(\theta|d) + \big(1-p_\text{astro}\big) \pi(\theta).
\end{align}

Eq.~\ref{eq:p_astro_theta} is an intuitive equation, which is highlighted by considering different limiting cases.
If the signal evidence is much larger than the noise evidence, the $\xi$-marginalized posterior reproduces the signal-hypothesis posterior:
\begin{align}
    \lim_{p_\text{astro}\rightarrow 1} 
    P_\text{astro}(\theta|d) = P(\theta|d) .
\end{align}
On the other hand if the noise evidence is much larger than the signal evidence, the $\xi$-marginalized posterior reproduces the prior
\begin{align}
    \lim_{p_\text{astro}\rightarrow0} 
    P_\text{astro}(\theta|d) = \pi(\theta) .
\end{align}
In between these two limiting cases, the posterior is a weighted average of the signal-hypothesis posterior and the prior.

There are practical consequences of this result for marginal events.
Consider the case where we observe an extraordinary event, which---if real---has important implications for astrophysics.
However, the hypothetical event is ambiguous.
We may choose to ask an astrophysical question of this  event such as: ``What is the probability of this event occurring in a mass gap?''
If there is any doubt as to whether the event in question is real, then this formalism provides a way to answer these questions in a statistically rigorous way.
It provides a more satisfying answer than a conditional answer, e.g., ``{\em If} the event is real, then the probability that the event is in the mass gap is...''

\subsection{Population studies with marginal events}\label{populations}
When there are multiple events, the astro likelihood becomes
\begin{align}
    {\cal L}_\text{astro}(\vec{d}|\Lambda) = 
    \prod_i^N \int d\theta_i \,
    {\cal L}_\text{astro}(d_i|\theta_i)
    \pi(\theta_i | \Lambda) .
\end{align}
The variable $\Lambda$ refers to population hyper-parameters.
Ignoring selection effects for the moment, the total astro likelihood is
\begin{align}\label{eq:Lastro_pop}
    {\cal L}_\text{astro}& (\vec{d}|\Lambda) = 
    & \xi^N \prod_i^N 
    \bigg(
    \frac{1-p_\text{astro}^i(\Lambda)}{p_\text{astro}^i(\Lambda)} 
    + 1 \bigg) 
    {\cal L}_\theta(d_i|\Lambda) ,
\end{align}
where
\begin{align}\label{eq:marg_likelihood}
    {\cal L}_\theta(d_i|\Lambda) = \int d\theta_i
    \, {\cal L}(d_i|\theta_i) \pi(\theta_i|\Lambda) .
\end{align}

Once again, it is instructive to consider the limiting cases for an event with different values of $p_\text{astro}$.
When the event is unambiguously astrophysical, the contribution to the likelihood becomes
\begin{align}
    \lim_{p_\text{astro}\rightarrow 1} 
   \bigg(
    \frac{1-p_\text{astro}^i(\Lambda)}{p_\text{astro}^i(\Lambda)} 
    + 1 \bigg) 
    {\cal L}_\theta(d_i|\Lambda)
    = {\cal L}_\theta(d_i|\Lambda) ,
\end{align}
which is the solution obtained assuming the event is definitely astrophysical.
On the other hand, if an event is of terrestrial origin, the contribution to the likelihood becomes
\begin{align}\label{eq:RecoverEffectiveNoise}
    \lim_{p_\text{astro}\rightarrow 0} 
    & \bigg(
    \frac{1-p_\text{astro}^i(\Lambda)}{p_\text{astro}^i(\Lambda)} 
    + 1 \bigg) 
    {\cal L}_\theta(d_i|\Lambda)
   \nonumber\\
    = & \frac{1-p_\text{astro}^i(\Lambda)}{p_\text{astro}^i(\Lambda)} {\cal L}_\theta(d_i|\Lambda) \nonumber\\
    \propto & {\cal L}_\text{eff}(d|\O) ,
\end{align}
which does not depend on $\Lambda$ (see Eq.~\ref{eq:effective_noise}), and thus does not influence our inferences about population hyper-parameters.
In the remainder of this paper, we write ${\cal L}(d|\Lambda)$ with no $\theta$ subscript for compact notation, though, it is understood that we have marginalized over $\theta$.

\subsection{Selection effects}\label{selection}
When we include selection effects, the signal likelihood changes:
\begin{align}\label{eq:marg_pdet_likelihood}
    {\cal L}(d|\Lambda,\text{det}) = & 
      \frac{1}{p_\text{det}(\Lambda)}
    {\cal L}(d|\Lambda) \nonumber \\
    = & \left(\frac{{\cal V}_\text{tot}}{{\cal V}(\Lambda)}\right)
    {\cal L}(d|\Lambda) .
\end{align}
The normalization factor ensures that the likelihood is correctly normalized following our decision to focus on data that includes a detection, denoted ``det''~\footnote{For additional details, see~\cite{intro}}.
Here, $p_\text{det}(\Lambda)$ is the probability of detecting an event drawn from the population described by $\Lambda$ while ${\cal V}(\Lambda)$ is the visible spacetime volume for which events are detected above some threshold:
\begin{align}
    {\cal V}(\Lambda) = \int d\theta \, {\cal V}(\theta)\pi(\theta | \Lambda).
\end{align}
The factor of ${\cal V}_\text{tot}$ is the total spacetime volume implied by the maximum comoving distance allowed by our prior distributions:
\begin{align}
    {\cal V}_\text{tot} = T_\text{obs} \int_0^{z_\text{max}} \frac{dV_c}{dz}\frac{1}{(1 + z)} dz .
\end{align}
Here, $V_c$ is the comoving volume, $T_\text{obs}$ is the total observation time, $z$ is redshift, and $z_\text{max}$ is the redshift corresponding to the maximum comoving distance.

Returning to our mixture model, we have
\begin{align}
    {\cal L}_\text{astro}(d|\Lambda,\text{det}) = & \xi {\cal L}(d|\Lambda,\text{det}) + (1-\xi){\cal L}(d|\O,\text{det})
    \nonumber\\
    = & \xi \frac{{\cal V}_\text{tot}}{{\cal V}(\Lambda)} {\cal L}(d|\Lambda) + \frac{1}{p_{\O}} (1-\xi){\cal L}(d|\O) .
\end{align}
Here, $p_{\O}$ is the normalization factor introduced by throwing out data that does not pass the detection criterion; we derive an expression for it below.
The definition of $p_\text{astro}$ does not depend on the detection threshold; we can calculate it for events with arbitrarily low signal-to-noise ratio.
Thus, the relationship between the $p_\text{astro}$ and the effective noise evidence is the same as it was before and so the astro likelihood becomes
\begin{align}\label{eq:single_event}
    {\cal L}_\text{astro} & (d|\Lambda,\text{det}) = \nonumber \\
    & \xi \Bigg( \frac{{\cal V}_\text{tot}}{{\cal V}(\Lambda)}  + \frac{1}{p_{\O}} 
    \left(\frac{1-p_\text{astro}(\Lambda)}{p_\text{astro}(\Lambda)}\right) 
    \Bigg) {\cal L}(d|\Lambda) .
\end{align}
We see that
\begin{align}
    {\cal L}_\text{astro}(d|\Lambda,\text{det}) \not\propto 
    \frac{1}{V(\Lambda)}
\end{align}
as one might naively expect.

It is difficult to determine $p_{\O}$ from first principles because of the complicated process by which data are matched filtered and detections are identified.
An alternative approach is to choose a value, which yields the correct behaviour for ${\cal L}_\text{astro}$.
We therefore consider the case where $p_\text{astro}=0.5$ so that the astrophysical hypothesis and the terrestrial hypothesis are given equal weight:
\begin{align}
    {\cal L}_\text{astro} & (d|\Lambda,\text{det}) = \xi \Bigg( \frac{{\cal V}_\text{tot}}{{\cal V}(\Lambda)} + \frac{1}{p_{\O}}
    \Bigg) {\cal L}(d|\Lambda) .
\end{align}
Next, we investigate how the likelihood varies for small perturbations around $\Lambda_0$, a fiducial first-guess for the population hyper-parameter, used to calculate preliminary significance estimates $p_\text{astro}(\Lambda_0)$.
In order to give equal weight to the signal and noise hypotheses in the vicinity of $\Lambda_0$, one finds
\begin{align}
    p_{\O} = \frac{{\cal V}(\Lambda_0)}{{\cal V}_\text{tot}} , 
\end{align}
where ${\cal V}(\Lambda_0)$ is the visible spacetime volume described by the fiducial model used to calculate $p_\text{astro}$.
We therefore adopt this value of $p_{\O}$ so that the astro likelihood is
\begin{align}\label{eq:marg_astro_likelihood}
    {\cal L}_\text{astro} & (d|\Lambda,\text{det}) \nonumber \\
    = & \xi \Bigg( \frac{{\cal V}_\text{tot}}{{\cal V}(\Lambda)} + \frac{{\cal V}_\text{tot}}{{\cal V}(\Lambda_0)} 
    \left(\frac{1-p_\text{astro}(\Lambda)}{p_\text{astro}(\Lambda)}\right) 
    \Bigg) {\cal L}(d|\Lambda) \nonumber\\
    = & \xi\frac{{\cal V}_\text{tot}}{{\cal V}(\Lambda_0)} \Bigg( \frac{{\cal V}(\Lambda_0)}{{\cal V}(\Lambda)}  +  
    \left(\frac{1-p_\text{astro}(\Lambda)}{p_\text{astro}(\Lambda)}\right)
    \Bigg) {\cal L}(d|\Lambda) .
\end{align}
Finally, writing our expression for $N$ events we obtain, 
\begin{align}\label{eq:Lastro_pop_det}
    {\cal L} & _\text{astro}  (\vec{d}|\Lambda,\text{det}) \nonumber \\
    & = \xi^N \left(\frac{{\cal V}_\text{tot}}{{\cal V}(\Lambda_0)}\right)^N 
    \prod_i^N \left(
    \frac{1-p_\text{astro}^i(\Lambda)}{p_\text{astro}^i(\Lambda)}
    + 
    {\frac{{\cal V}(\Lambda_0)}{{\cal V}(\Lambda)}}
    \right)
    {\cal L}(d_i|\Lambda).
\end{align}
Conveniently, the ${\cal V}_\text{tot}$ term becomes an overall multiplicative constant.
We use Eq.~\ref{eq:Lastro_pop_det} to derive the final likelihood for population analysis with selection effects and ambiguous detections.

\subsection{Poisson statistics and merger rates}\label{rates}
Next, following~\cite{tbs} (see their Section IIC), we promote $\xi$ to a hyper-parameter.
The prior on $\xi$, which is conditional on $N$, is related to a likelihood function:
\begin{align}
    \pi(\xi|N) = n {\cal L}_\text{Poisson}&(N|R,R_g,\Lambda) .
\end{align}
Here, $n=N/\xi$ is the number of analysis segments and ${\cal L}(N|...)$ is the likelihood of getting $N$ events given an astrophysical rate $R$ and a glitch rate $R_g$.

This likelihood of $N$ detections is Poisson distributed
\begin{align}
    {\cal L}_\text{Poisson}&(N|R,R_g,\Lambda) = \nonumber\\
    & \frac{\left(R{\cal V}(\Lambda) + R_g T_\text{obs}\right)^N}{ N!}
    e^{-\left(R{\cal V}(\Lambda) + R_g T_\text{obs}\right)} .
\end{align}
Thus, the ``total likelihood'' for both $N$ and $\vec{d}$ is
\begin{widetext}\begin{align}\label{eq:Lastro_rates}
    {\cal L}_\text{tot} (\vec{d},N|\Lambda, R, R_g, \text{det}) 
    = &
    \left(\frac{N}{n}\right)^{N-1}
    \frac{{\big(R{\cal V}(\Lambda) + R_g T_\text{obs}\big)^N}}{(N-1)!}
    e^{-\left(R{\cal V}(\Lambda) + R_g T_\text{obs}\right)} \nonumber\\
    & 
    \left(\frac{{\cal V}_\text{tot}}{{\cal V}(\Lambda_0)}\right)^N
    \prod_i^N \left(
    \frac{1-p_\text{astro}^i(\Lambda)}{p_\text{astro}^i(\Lambda)}
    + 
    {\frac{{\cal V}(\Lambda_0)}{{\cal V}(\Lambda)}}
    \right)
    {\cal L}(d_i|\Lambda) \nonumber\\
    \propto & 
    {\big(R{\cal V}(\Lambda) + R_g T_\text{obs}\big)^N}
    e^{-\left(R{\cal V}(\Lambda) + R_g T_\text{obs}\right)}
    \prod_i^N \left(
    \frac{1-p_\text{astro}^i(\Lambda)}{p_\text{astro}^i(\Lambda)}
    + 
    {\frac{{\cal V}(\Lambda_0)}{{\cal V}(\Lambda)}}
    \right)
    {\cal L}(d_i|\Lambda) .
\end{align}\end{widetext}
In the last line, we leave off multiplicative constants as they can be ignored for (hyper-) parameter estimation and model selection.

\subsection{Putting everything together}\label{everything_together}
All the ingredients are now in place.
What is left is for us to perform some final manipulations in order to write the total likelihood in the most useful form.
Demanding that the effective noise is independent of $\Lambda$, Eq.~\ref{eq:effective_noise} yields
\begin{align}\label{eq:no_Lambda_dependence}
    \left(\frac{1-p_\text{astro}(\Lambda)}{p_\text{astro}(\Lambda)}\right) &
    {\cal L}(d|\Lambda) = \left(\frac{1-p_\text{astro}(\Lambda_0)}{p_\text{astro}(\Lambda_0)}\right) 
    {\cal L}(d|\Lambda_0) .
\end{align}
We use Eq.~\ref{eq:no_Lambda_dependence} to rewrite Eq.~\ref{eq:Lastro_rates}, yielding
\begin{widetext}\begin{align}
    {\cal L}_\text{tot}  (\vec{d},N|\Lambda, & R, R_g, \text{det}) 
    \nonumber\\ 
    \propto & {\big(R{\cal V}(\Lambda) + R_g T_\text{obs}\big)^N}
    e^{-\left(R{\cal V}(\Lambda) + R_g T_\text{obs}\right)}  
    \prod_i^N \bigg(
    \frac{1-p_\text{astro}^i(\Lambda)}{p_\text{astro}^i(\Lambda)} 
    +  
    {\frac{{\cal V}(\Lambda_0)}{{\cal V}(\Lambda)}}
    \bigg) 
    {\cal L}(d_i|\Lambda) \nonumber\\
    \propto & 
    {\big(R{\cal V}(\Lambda) + R_g T_\text{obs}\big)^N}
    e^{-\left(R{\cal V}(\Lambda) + R_g T_\text{obs}\right)} 
    \prod_i^N \left(
    \frac{1-p_\text{astro}^i(\Lambda_0)}{p_\text{astro}^i(\Lambda_0)}
    +
    {\left(\frac{{\cal V}(\Lambda_0)}{{\cal V}(\Lambda)}\right)}
    \frac{{\cal L}(d_i|\Lambda)}{{\cal L}(d_i|\Lambda_0)}
    \right)
    {\cal L}(d_i|\Lambda_0) \nonumber\\
\end{align}\end{widetext}
This result is useful because we do not have to recalculate $p_\text{astro}$ in order to do population inference; we can use the $p_\text{astro}(\Lambda_0)$ published in transient catalogs.
Next, we make the usual approximation to rewrite the ratio of likelihoods as a sum that recycles posterior samples \cite{intro}:
\begin{widetext}\begin{align}\label{eq:Ltot}
    {\cal L}_\text{tot} & (\vec{d},N|\Lambda, R, R_g, \text{det}) 
    \propto \nonumber\\
    & {\big(R{\cal V}(\Lambda) + R_g T_\text{obs}\big)^N}
    e^{-\left(R{\cal V}(\Lambda) + R_g T_\text{obs}\right)} 
    \prod_i^N \left(
    \frac{1-p_\text{astro}^i(\Lambda_0)}{p_\text{astro}^i(\Lambda_0)}
    + 
    {\left(\frac{{\cal V}(\Lambda_0)}{{\cal V}(\Lambda)}\right)}
    \frac{1}{n_i}\sum_k^{n_i} \frac{\pi(\theta_{i,k}|\Lambda)}{\pi(\theta_{i,k}|\Lambda_0)} 
    \right)
    {\cal L}(d_i|\Lambda_0) .
\end{align}\end{widetext}
We marginalize over $R$ and $R_\text{g}$ using a Jeffrey's prior for both the merger rate and glitch rate,
\begin{align}
    \pi(R) \propto & \frac{1}{\sqrt{R}} ,
\end{align}
\begin{align}
    \pi(R_\text{g}) \propto & \frac{1}{\sqrt{R_\text{g}}} .
\end{align}
The result of the marginalization integral is the main punchline of this section:
\begin{widetext}\begin{equation}\label{eq:Upsilon}
    \boxed{
    \begin{aligned}
    {\cal L}_\text{tot}(\vec{d},N|\Lambda,R_g,\text{det})
    = & \int dR \, {\cal L}_\text{tot}(\vec{d},N|\Lambda, R, R_g, \text{det}) \pi(R) \pi(R_\text{g}) \\
    \propto & 
    {\frac{1}{{\cal V}(\Lambda)^{1/2}}}
    \Upsilon(R_g T, N)
    \prod_i^N \left(
    \frac{1-p_\text{astro}^i(\Lambda_0)}{p_\text{astro}^i(\Lambda_0)}
    + 
    {\left(\frac{{\cal V}(\Lambda_0)}{{\cal V}(\Lambda)}\right)}
    \frac{1}{n_i}\sum_k^{n_i} \frac{\pi(\theta_{i,k}|\Lambda)}{\pi(\theta_{i,k}|\Lambda_0)} 
    \right)
    {\cal L}(d_i|\Lambda_0) .
    \end{aligned}
    }
\end{equation}\end{widetext}
Here, $\Upsilon(R_g T, N)$ is some function for which we can obtain the analytic form using, e.g., {\sc Mathematica}, 
\begin{align}
    {\cal V}(\Lambda)&^{-1/2} \Upsilon(R_g T,N) \equiv \int dR
    \int dR_g \frac{1}{\sqrt{R_g R}}
    \nonumber\\
    & (R{\cal V}(\Lambda)+R_g T_\text{obs})^N 
    e^{-(R{\cal V}(\Lambda)+R_g T_\text{obs})}
\end{align}
The function $\Upsilon$ contains hyper-geometric functions, and goes to $1$ in the limit that $R_g=0$.
Conveniently, the $R_g$ dependence of the likelihood is entirely encoded in $\Upsilon$, which does not depend on $\Lambda$.
Thus, the likelihood factorizes, allowing us to effectively ignore $R_g$ when making inferences about $\Lambda$.
We obtain a familiar expression for ${\cal L}_\text{tot}$ when we take the limit that $R_g\rightarrow0$ and $p_\text{astro}\rightarrow1$.
In the limit that $p_\text{astro}\rightarrow0$, we get the correct scaling with ${\cal V}(\Lambda)$; i.e., we recover our prior.
Following hyper-parameter estimation of $\Lambda$, posterior distributions for $R$ and $R_g$ may be reconstructed in post-processing using Eqs.~\ref{eq:Ltot} and ~\ref{eq:Upsilon} by inverse transform sampling from the analytic CDF.

\subsection{Updating $p_\text{astro}$}\label{update}
Returning to Eq.~\ref{eq:no_Lambda_dependence}, we can calculate $p_\text{astro}(\Lambda)$
\begin{align}\label{eq:calculating_pastro}
    p_\text{astro}(\Lambda) = &
    \left[
    \left(\frac{1 - p_{\text{astro}}(\Lambda_{0})}
    {p_{\text{astro}}(\Lambda_{0})}\right)
    \frac{{\cal L}(d|\Lambda_{0})}
    {{\cal L}(d|\Lambda)}
    + 1\right]^{-1} ,
\end{align}
which is the revised astrophysical probability in light of what we have learned about the distribution of black hole mass and spin from the greater catalog.
This equation passes the sanity check that $p_\text{astro}(\Lambda_0)=p_\text{astro}(\Lambda_0)$ when we set $\Lambda=\Lambda_0$.

This method of updating $p_\text{astro}$ can be used for event classification as in~\cite{pastro}.
Following~\cite{pastro}, one can define hyper-parameters that correspond to different categories of events, for example: binary black holes ($\Lambda_\text{BBH}$), binary neutron stars ($\Lambda_\text{BNS}$), and neutron star-black hole binaries ($\Lambda_\text{NSBH}$).
One can calculate relative probabilities for different astrophysical categories, for example,
\begin{align}
    p_\text{BBH} =
    \frac{p_\text{astro}(\Lambda_\text{BBH})}
    {p_\text{astro}(\Lambda_\text{BNS}) + 
    p_\text{astro}(\Lambda_\text{BBH}) + 
    p_\text{astro}(\Lambda_\text{NSBH})} .
\end{align}

\section{Demonstration}\label{demonstration}
\subsection{Overview}\label{overview}
In this Section, we demonstrate our formalism using 10 events from GWTC-1~\cite{gwtc1} and 8 from the IAS catalog~\cite{ias2,ias3,ias4} using LIGO/Virgo open data~\cite{gwosc}. 
Our analyses include events with  $p_\text{astro}>0.5$.
We exclude the IAS event GW170402, which is described as only a candidate event. 
A complete list of the events and their respective $p_{\text{astro}}$ values are provided in Tab.~\ref{tab:ias}. 
In subsection~\ref{single-event}, we present posterior distributions for the marginal ($p_\text{astro}=0.71$) IAS event GW151216.
We highlight how inferences about this event change when we take into account its uncertain origin.
In subsection~\ref{updated-catalog}, we carry out a GWTC-1 population study based on work in~\cite{ratepopO2} in order to investigate how our results vary depending on how we handle the origin of ambiguous events.
In subsection~\ref{ias} we investigate how these results change when we include events from IAS.
The calculations in this Section are carried out using the \textsc{Bilby} \cite{bilby} with the nested sampler \textsc{dynesty} \cite{dynesty}.
For our population analyses we use the models from \cite{Talbot2018}, as implemented in the \texttt{GWPopulation} package \cite{Talbot2019}. 
For our initial, single-event parameter estimation, we sample in priors which are uniform in component masses, uniform in dimensionless spin, and isotropic in spin orientation. We then re-weight our samples to priors in uniform component masses.
We assume standard priors for extrinsic parameters.

\begin{table}[h!]
\begin{tabular}{l c c c r} 
    \hline
    Event & $m_1^\text{source}$ ($M_\odot$)& $\chi_\text{eff}$ & $p_\text{astro}$ & Catalog  \\ 
    \hline\hline
    GW150914 & [32.50, 39.81] & [-0.15, 0.07] & 1.00 & GWTC-1 \cite{gwtc1} \\
    GW151012 & [17.88, 35.50] & [-0.15, 0.27] & 0.96 & GWTC-1 \cite{gwtc1} \\
    GW151216 & [19.94, 50.01] & [-0.21, 0.56] & 0.71 & IAS \cite{ias2} \\
    GW151226 & [10.61, 20.09] & [0.12, 0.35] & 1.00 & GWTC-1 \cite{gwtc1} \\
    GW170104 & [25.59, 38.74] & [-0.23, 0.12] & 1.00 & GWTC-1 \cite{gwtc1} \\
    GW170121 & [27.64, 41.06] & [-0.44, -0.00] & 1.00 & IAS \cite{ias3} \\
    GW170202 & [22.75, 44.18] & [-0.36, 0.12] & 0.68 & IAS \cite{ias3} \\
    GW170304 & [35.69, 56.00] & [-0.12, 0.34] & 0.985 & IAS \cite{ias3} \\
    GW170403 & [39.59, 70.60] & [-0.48, 0.10] & 0.56 & IAS \cite{ias3} \\
    GW170425 & [34.46, 63.26] & [-0.29, 0.22] & 0.77 & IAS \cite{ias3} \\
    GW170608 & [9.28, 15.46] & [-0.01, 0.19] & 1.00 & GWTC-1 \cite{gwtc1} \\
    GW170727 & [33.62, 51.55] & [-0.28, 0.16] & 0.98 & IAS \cite{ias3} \\
    GW170729 & [41.99, 72.86] & [0.02, 0.47] & 0.52 & GWTC-1 \cite{gwtc1} \\
    GW170809 & [29.04, 43.59] & [-0.10, 0.22] & 1.00 & GWTC-1 \cite{gwtc1} \\
    GW170814 & [27.54, 35.42] & [-0.05, 0.19] & 1.00 & GWTC-1 \cite{gwtc1} \\
    GW170817A & [52.04, 84.02] & [0.01, 0.46] & 0.86 & IAS \cite{ias4} \\
    GW170818 & [30.80, 42.72] & [-0.29, 0.10] & 1.00 & GWTC-1 \cite{gwtc1} \\
    GW170823 & [32.88, 50.52] & [-0.16, 0.23] & 1.00 & GWTC-1 \cite{gwtc1} \\
    \hline
\end{tabular}
\caption{The gravitational-wave events used in this analysis. We list the 90\% credible interval for $m_1^\text{source}$, the primary source-frame mass of the compact binary and $\chi_\text{eff}$, the effective aligned spin of the binary---obtained here with \textsc{Bilby}. These credible intervals are obtained using the posterior distributions not weighted by $p_\text{astro}$. We also include $p_\text{astro}$, and the catalog that provides the listed value of $p_\text{astro}$.}
\label{tab:ias}
\end{table}

Recent LIGO/Virgo publications have listed up to three different values for $p_\text{astro}$, corresponding to the values obtained using three different detection pipelines~\cite{gwtc1}.
For some events, the numerical value of $p_\text{astro}$ can vary greatly.
Recent detection claims by IAS have yielded yet more variability in $p_\text{astro}$.
We agree with the authors of~\cite{bcr2} that this is problematic.
In our view, the origin of an event should be independent of the search pipeline used to first identify it.
While preliminary work has been undertaken to produce a single, pipeline-independent $p_\text{astro}$~\cite{bcr2}, there is, at present, no published catalog of pipeline-independent $p_\text{astro}$.

Thus, as a temporary measure, we use $p_\text{astro}$ values from the \textsc{PyCBC}~\cite{pycbc} pipeline for GWTC-1, except for GW170818 for which no \textsc{PyCBC} value is available, and so we use the $p_\text{astro}$ value from the \textsc{GstLAL}-based inspiral pipeline~\cite{gstlal}.
For events that are part of the IAS catalog, but which are not included in GWTC-1, we take the IAS value of $p_\text{astro}$ at face value.
Two of the GWTC-1 events have $p_\text{astro}$ values measurably different from unity: GW170729 ($p_{\text{astro}}=0.52$) and GW151012 ($p_{\text{astro}}=0.96$).
All of the IAS events have $p_\text{astro}$ values measured to be less than unity except for GW170121.

\subsection{GW151216: an ambiguous event}\label{single-event}
\begin{figure*}
    \begin{subfigure}{\columnwidth}
    \includegraphics[width=\linewidth]{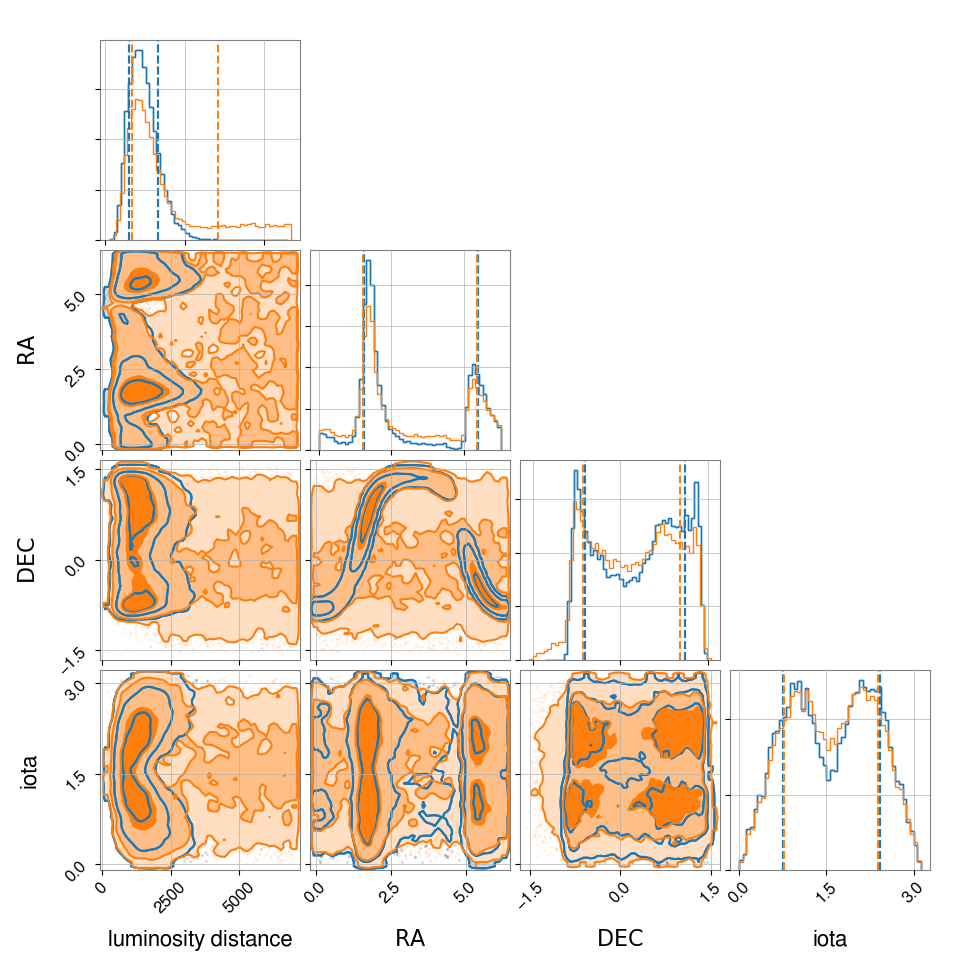} 
    \end{subfigure}
    \begin{subfigure}{\columnwidth}
    \includegraphics[width=\linewidth]{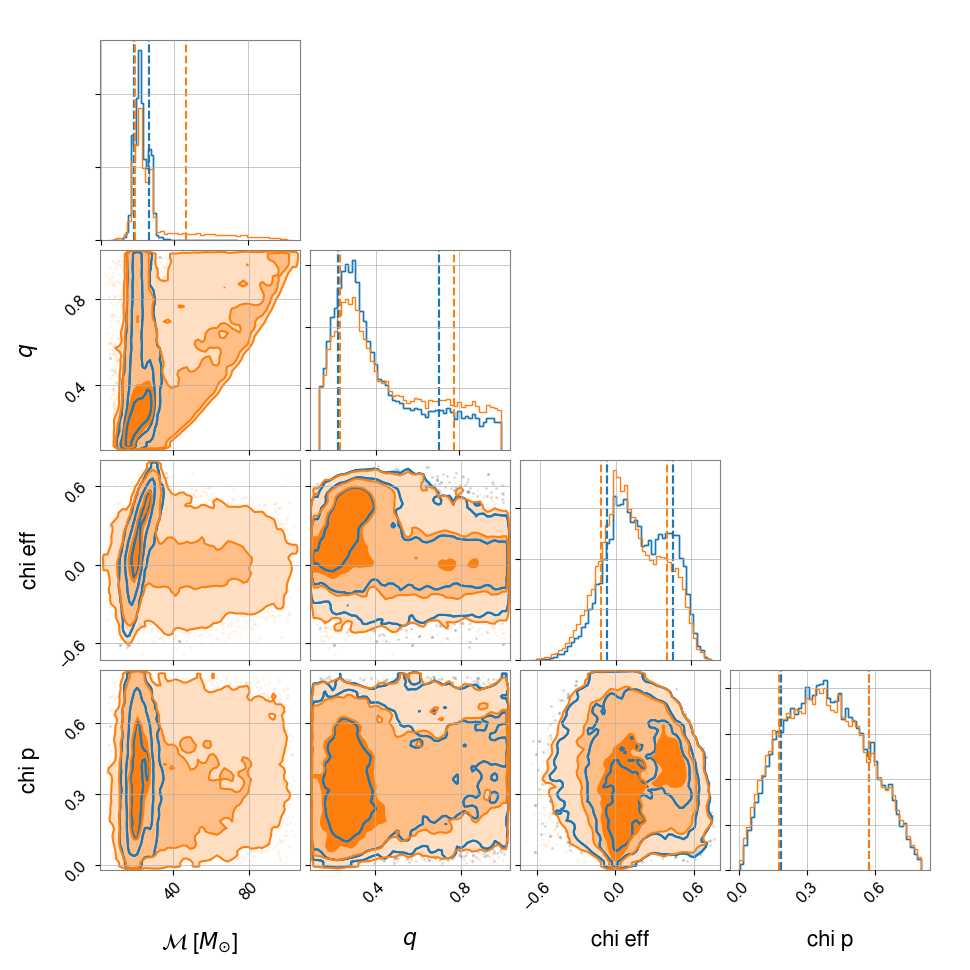}
    \end{subfigure}
    \caption{Posterior distribution of extrinsic (left) and intrinsic (right) parameters for the IAS event GW151216 with $p_{\text{astro}}=0.71$. The blue contours represent the posterior distributions given the astrophysical hypothesis while the orange contours allow for the possibility of terrestrial origin.}
    \label{fig:GW151216}
\end{figure*}

In order to illustrate the results from subsection~\ref{single-event}, we derive posterior distributions for GW151216, an IAS event with $p_\text{astro}=0.71$.
The results are shown in Fig.~\ref{fig:GW151216}.
In this figure and subsequent figures throughout this work, we indicate credible intervals at $1\sigma$, $2\sigma$ and $3\sigma$, using increasingly light shading.
In blue we plot the posterior assuming the event is definitely of astrophysical origin while in orange we plot the ``astro'' posterior (Eq.~\ref{eq:p_astro_theta}), which allows for the possibility of a terrestrial origin.
As expected, the posterior distributions widen considerably when we take into account the possibility of terrestrial origin.
Similar posterior plots for the other seven IAS events are included in Appendix~\ref{sec:appendix-singleevent}.

\subsection{Population studies with GWTC-1}\label{updated-catalog}
We employ the methodology from Section~\ref{formalism} to measure the mass and spin distributions of binary black holes.
We employ black hole mass ``Model C''from~\cite{ratepopO2}.
For black hole spins, we employ the ``Gaussian'' model. 
The mass model consists of a power-law that tapers off at both ends of the mass function and has a Gaussian component at the upper end of the mass distribution. 
The mass distribution is modeled by eight hyper-parameters: $\alpha$, the spectral index of the primary mass power-law distribution; $\beta$, the spectral index of the mass ratio distribution; $\delta m$, the mass range over which the low-mass part of the black hole mass spectrum tapers off;  $m_\text{min}$, the minimum mass of the distribution; and $m_\text{max}$, the maximum mass of the distribution. 
The Gaussian high-mass peak is described by three hyper-parameters: $\lambda$, the fraction of black holes in the peak; $\mu_\text{m}$, the mean mass of the peak; and $\sigma_\text{m}$, the standard deviation of the peak.
For additional details, see~\cite{ratepopO2,Talbot2018}.

The spin orientation is described using a truncated Gaussian as in~\cite{Talbot2017}. 
Meanwhile, the dimensionless spin magnitude is described by a beta function as in~\cite{Wysocki2018}.
The spin model is described by six parameters: $\zeta$, the fraction of binaries with isotropic spin orientations; $a_\text{max}$, the maximum spin; $\mu_{\chi}$, the mean value of the distribution of spin magnitudes; $\sigma^2_{\chi}$, the variance of the distribution of spin magnitudes; and $\sigma_{t}$, the width of the distribution of spin orientations for the preferentially aligned spin component. 
We use a fixed value of $\zeta=1$ and $a_\text{max}=0.8$ for our analysis. 
More detail on these models can be found in~\cite{Wysocki2018,Talbot2017,ratepopO2}.

\begin{figure*}
    \begin{subfigure}{\columnwidth}
    \includegraphics[width=\linewidth]{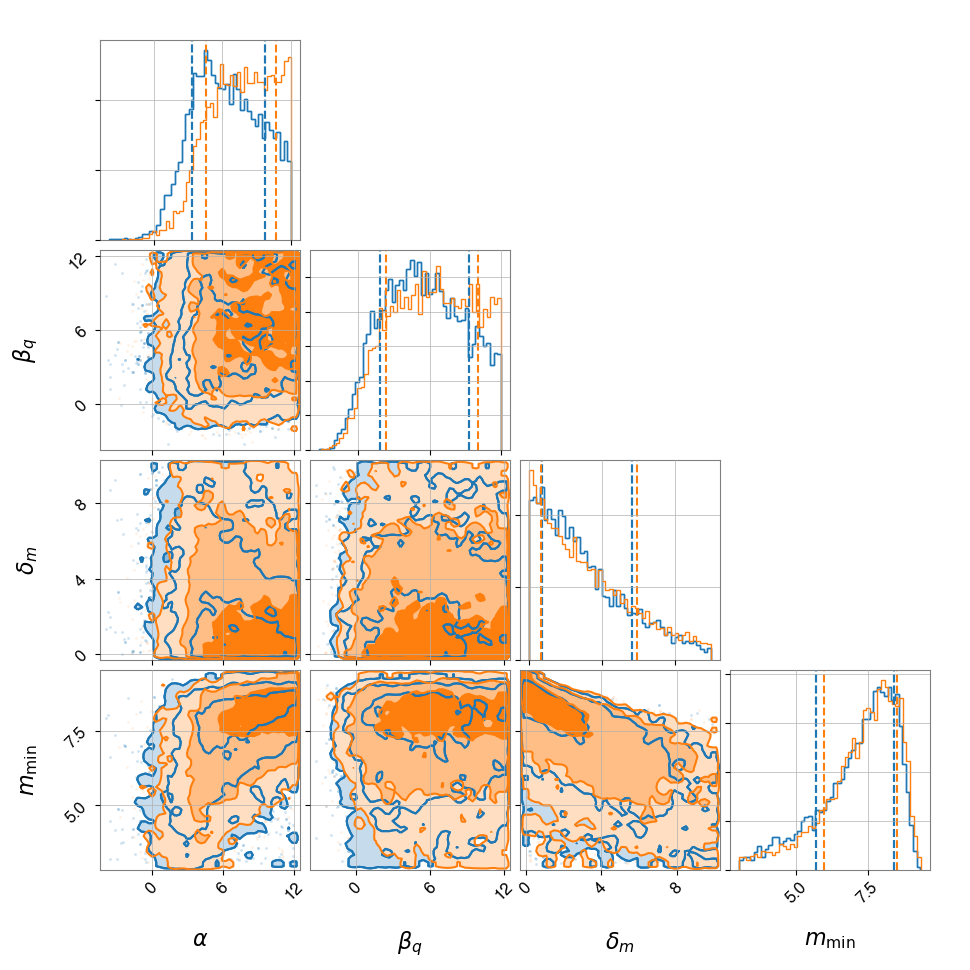} 
    \end{subfigure}
    \begin{subfigure}{\columnwidth}
    \includegraphics[width=\linewidth]{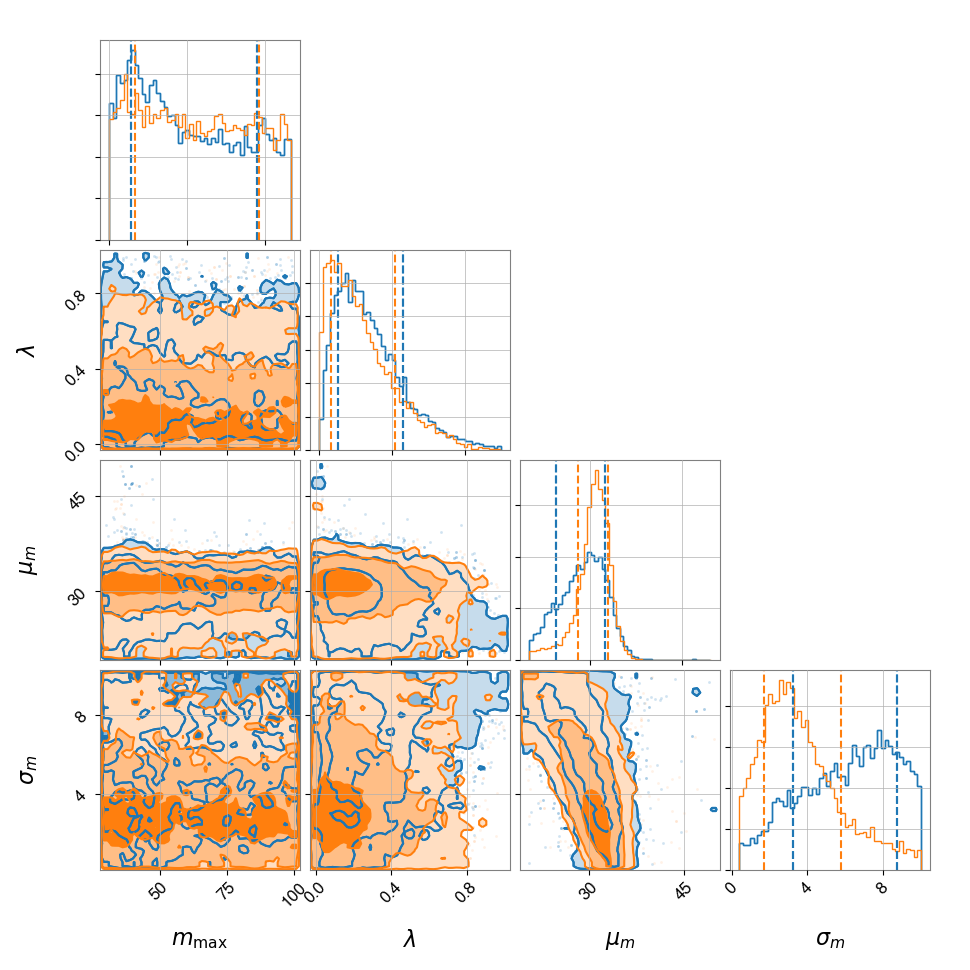}
    \end{subfigure}
    \caption{Population posterior distributions of GWTC-1 catalog events for the mass hyper-parameters associated with the lower (left) and upper (right) mass peaks in the primary mass distribution. The blue contours are the posterior distributions without taking into account $p_\text{astro}$, and the orange contours are the posterior distributions with $p_{\text{astro}}$ included.}
    \label{fig:gwtc1_pop_mass}
\end{figure*}

Figure~\ref{fig:gwtc1_pop_mass} shows the posterior distribution for the mass hyper-parameters using only the GWTC-1 events. 
The blue contours are obtained weighting every event equally ($p_\text{astro}=1$) while the orange contours take into account the relative origin of each event as per Eq.~\ref{eq:Upsilon}.
Including information about $p_\text{astro}$ reduces support for a deviation from a power-law distribution (pushing $\lambda$ toward zero) while improving our estimate of the mean and variance of the putative pulsational pair-instability graveyard (producing more peaked distributions of $\mu_m$ and $\sigma_m$ respectively).
This change can be explained by GW170729, which has the lowest $p_\text{astro}$ value in GWTC-1, but which includes support for higher primary mass than any other event.
If GW170729 is of astrophysical origin, it should fall within the Gaussian component, but this requires the peak to be broader.

\begin{figure*}
    \includegraphics[width=0.9\columnwidth]{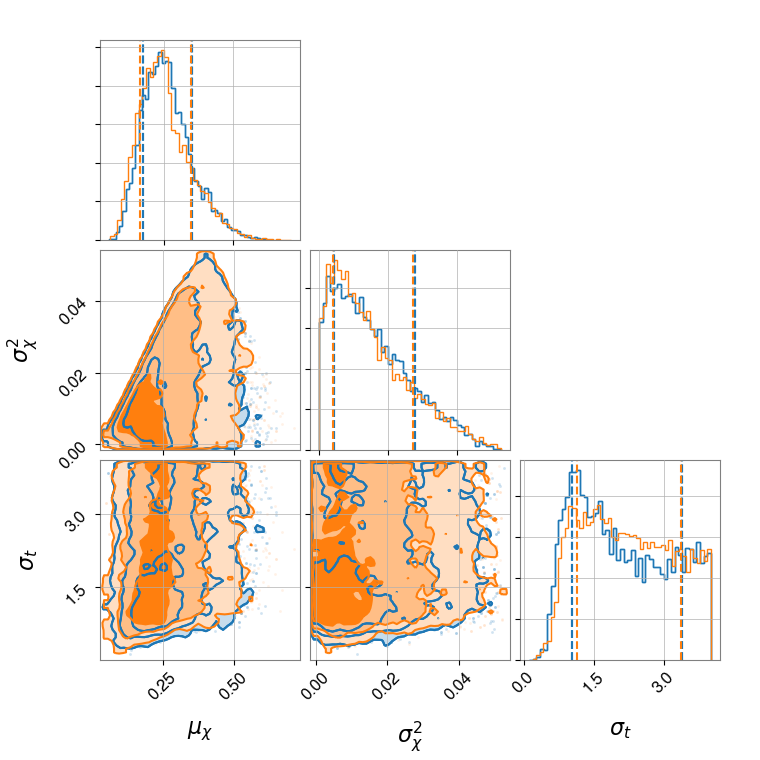}
    \caption{Population posterior distributions of GWTC-1 catalog events for the spin hyper-parameters. The blue contours are the posterior distributions without taking into account $p_\text{astro}$, and the orange contours are the posterior distributions with $p_{\text{astro}}$ included.}
    \label{fig:gwtc1_pop_spin}
\end{figure*}

Figure~\ref{fig:gwtc1_pop_spin} shows the posterior distributions for the spin hyper-parameters using only events from GWTC-1. 
As before, orange includes $p_\text{astro}$ weighting while blue does not.
The quite modest differences between blue and orange indicate that GW170729 does not provide a great deal of information about the spin distribution of binary black holes.
The most noticeable change is a slight preference for in-plane spin, indicated by a slight shift in the $\sigma_t$ posterior.

\subsection{Population studies with GWTC-1 and the IAS catalog}\label{ias}
\begin{figure*}
    \begin{subfigure}{\columnwidth}
    \includegraphics[width=\linewidth]{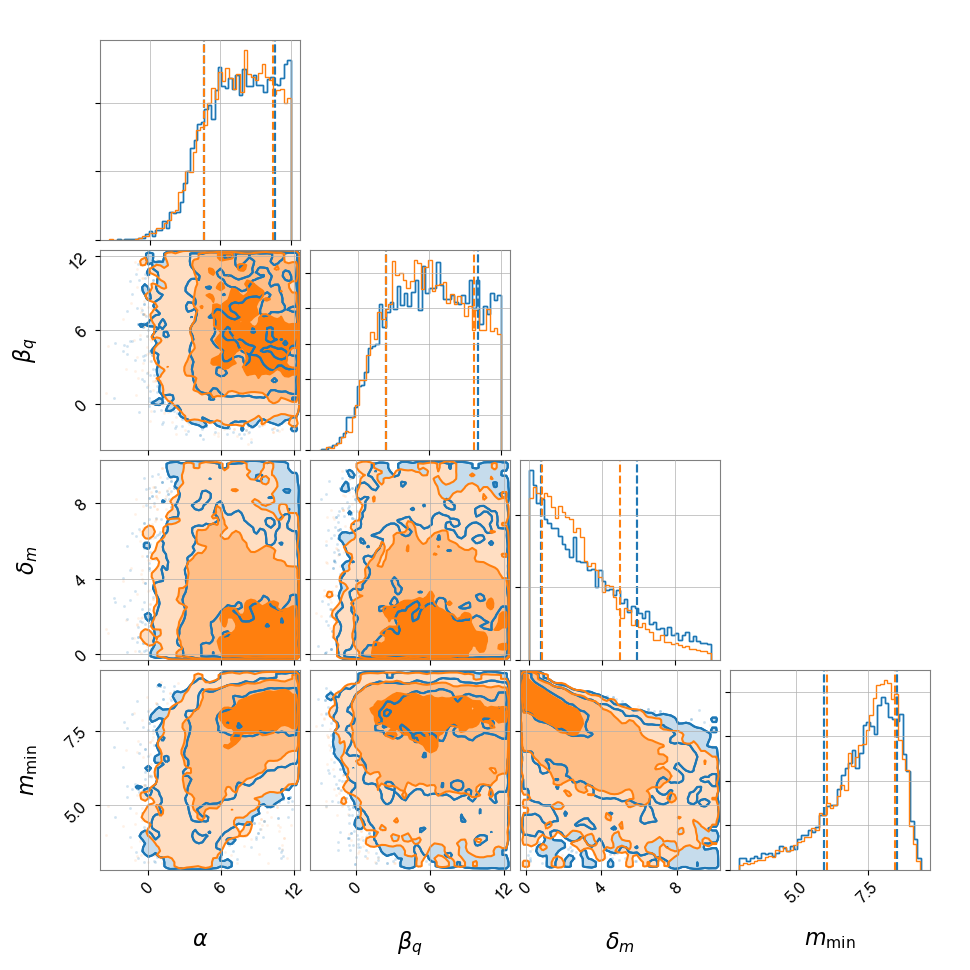} 
    \end{subfigure}
    \begin{subfigure}{\columnwidth}
    \includegraphics[width=\linewidth]{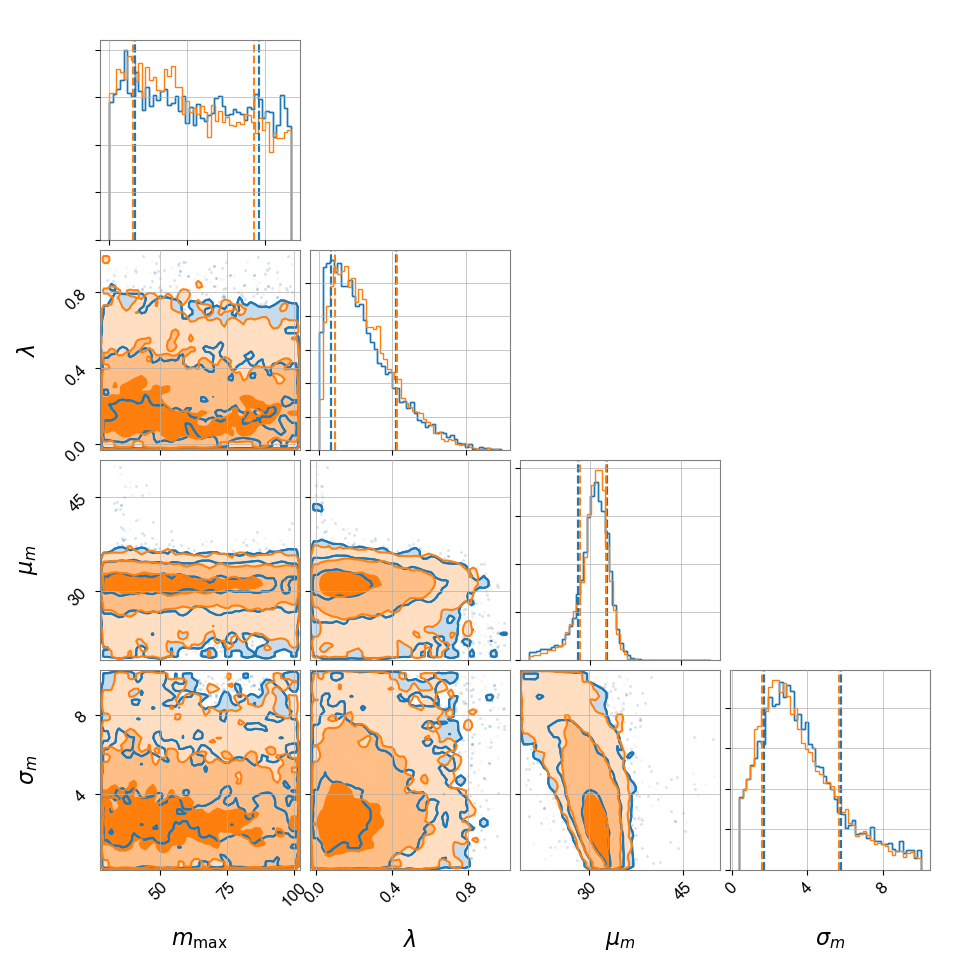}
    \end{subfigure}
    \caption{Population posterior distributions of GWTC-1 and IAS catalog events for the mass hyper-parameters associated with the lower (left) and upper (right) mass peaks in the primary mass distribution. The blue contours are the posterior distributions for the GWTC-1 events only with $p_{\text{astro}}$ included, and the orange contours are the posterior distributions of the GWTC-1 and IAS events with $p_{\text{astro}}$ included.}
    \label{fig:gwtc1_ias_pop_mass}
\end{figure*}

\begin{figure*}
    \includegraphics[width=0.9\columnwidth]{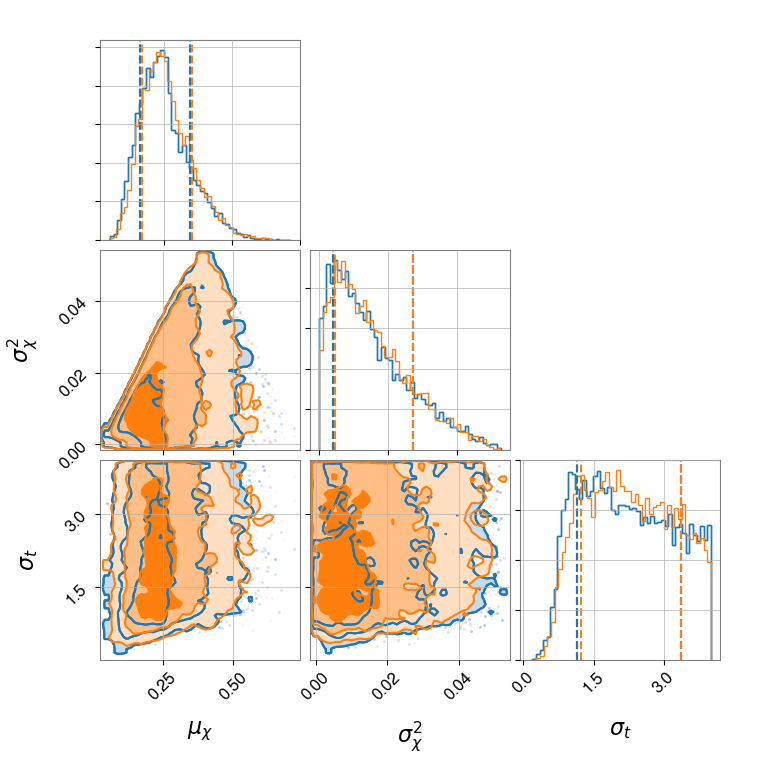}
    \caption{Population posterior distributions of GWTC-1 and IAS catalog events for the spin hyper-parameters; the blue contours represent the posterior distributions for the GWTC-1 events only with $p_{\text{astro}}$ included, and the orange contours represent the posterior distributions of the GWTC-1 and IAS events with $p_{\text{astro}}$ included}
    \label{fig:gwtc1_ias_pop_spin}
\end{figure*}

In Fig.~\ref{fig:gwtc1_ias_pop_mass}, we show the posterior distributions for the mass hyper-parameters using events from both GWTC-1 and IAS.
Meanwhile, in Fig.~\ref{fig:gwtc1_ias_pop_spin}, we show the posterior distributions for the spin hyper-parameters using events from both GWTC-1 and IAS.
Blue indicates the results with GWTC-1 only while orange shows how the results change with the inclusion of IAS.
Both results use the $p_\text{astro}$ weighting from Eq.~\ref{eq:Upsilon}.
We observe only small differences between blue and orange contours, indicating that the inclusion of the IAS events does not provide much resolving power beyond what is achieved with GWTC-1 alone.
This is somewhat surprising as the eight IAS events contribute $\sum_i p_\text{astro}^i = 6.5$ effective events, which constitutes a non-negligible increase in the number of events compared to GWTC-1 alone.

\begin{figure*}
    \includegraphics[width=\linewidth]{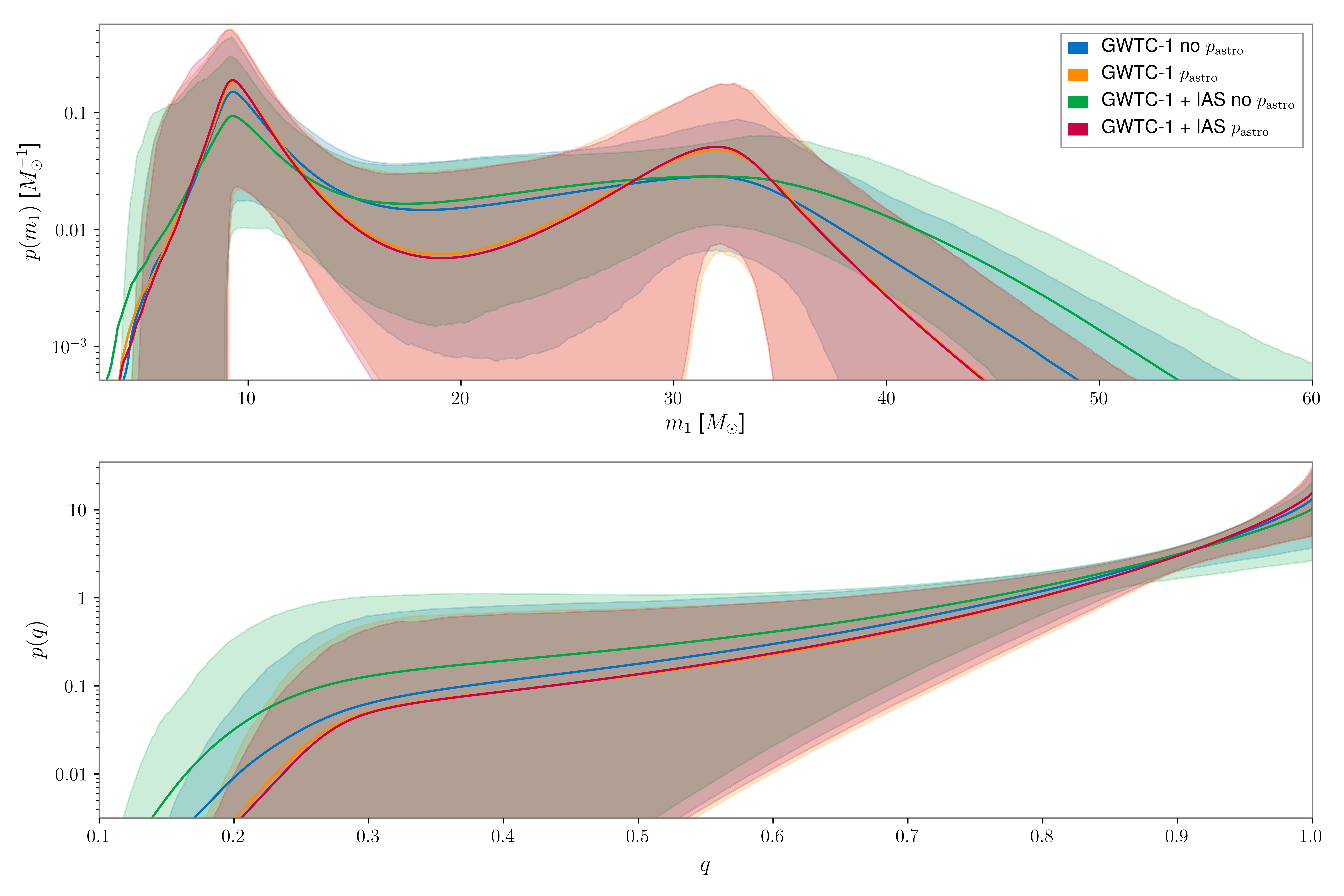}
    \caption{Reconstructed distributions for primary mass ($m_1$) and mass ratio ($q$). Blue is the GWTC-1 catalog without $p_\text{astro}$; orange is the GWTC-1 catalog with $p_\text{astro}$; green is GWTC-1 and IAS catalogs without $p_\text{astro}$; and red is GWTC-1 and IAS catalogs with $p_\text{astro}$. The solid curves indicate the posterior predictive distributions, and the shaded regions represent the 90\% credible region.}
    \label{fig:mass_spectrum}
\end{figure*}

\begin{figure*}
    \includegraphics[width=\linewidth]{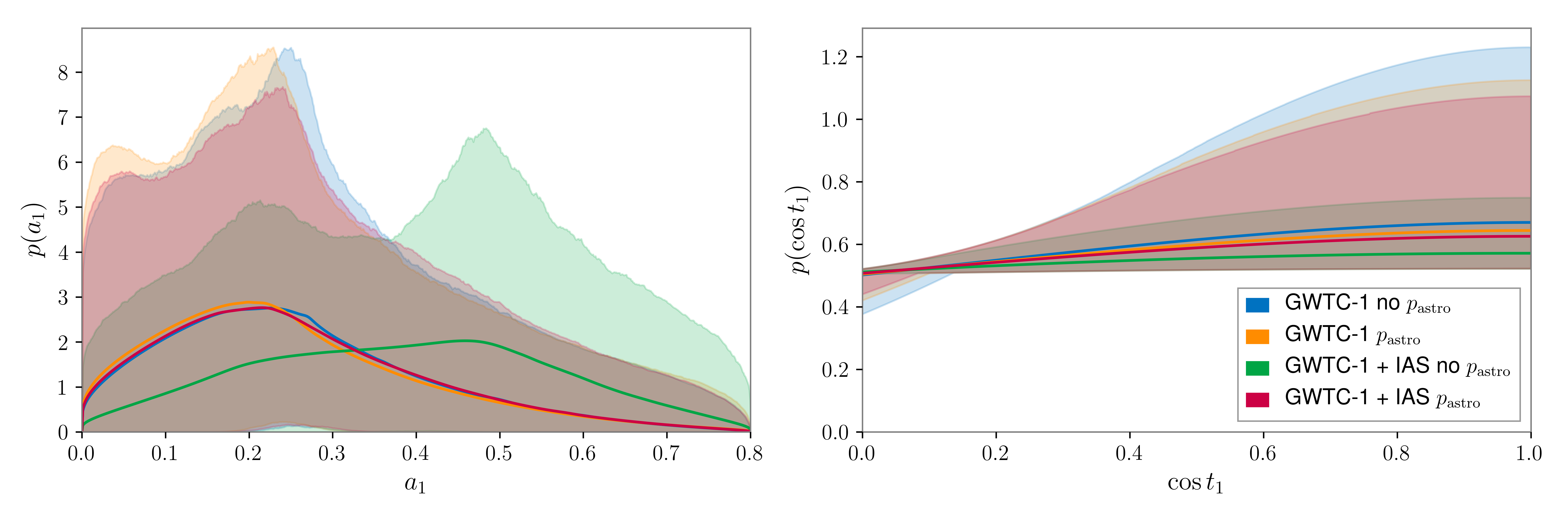}
    \caption{Reconstructed distributions for spin magnitude ($a_1$) and orientation ($\cos t_{1}$). Blue is for the GWTC-1 catalog without $p_\text{astro}$; orange is GWTC-1 catalog with $p_\text{astro}$; green is GWTC-1 and IAS catalogs without $p_\text{astro}$; and red is and GWTC-1 and IAS catalogs with $p_\text{astro}$. The solid curves indicate the posterior predictive distributions, and the shaded regions represent the 90\% credible region.}
    \label{fig:spin_magnitude_orientation_spectrum}
\end{figure*}

To further highlight the difference between the distributions of the mass and spin parameters informed by our population, we plot the reconstructed mass spectrum (Fig.~\ref{fig:mass_spectrum}) and spin spectrum (Fig.~\ref{fig:spin_magnitude_orientation_spectrum}) for the population distributions for four population studies.
In blue we plot GWTC-1 without accounting for $p_\text{astro}$; in orange we plot GWTC-1 weighted by $p_\text{astro}$; in green we plot GWTC-1 and IAS events without accounting for $p_\text{astro}$; and in red we plot GWTC-1 and IAS events weighted by $p_\text{astro}$. 
The solid curves are posterior predictive distributions while the shaded region indicates the 90\% credible interval.
If we do not account for $p_\text{astro}$, the inclusion of the IAS events pulls the posterior predictive distribution to higher masses, lower mass ratios, and higher spin magnitudes.

\section{Catalog update}\label{reweight}
In this Section, we use the results of the population study presented in the previous section to provide updates to GWTC-1 and IAS catalogs.
First, in Subsection~\ref{leave-one-out}, we use the mass and spin distributions inferred above to create an astrophysically motivated prior, which we use to reanalyze each event.
By including this contextual information, it is possible to provide improved constraints on the parameters of individual events.
Second, in Subsection~\ref{pastro-update}, we calculate updated values of $p_\text{astro}$, taking into account what we have learned about the population properties of black holes from GWTC-1 and the IAS catalog.

\subsection{Reanalysis of GWTC-1 with an astrophysically-motivated prior}\label{leave-one-out}
Often, Bayesian parameter estimation for individual gravitational-wave events is carried out using priors that are uniform in component masses, uniform in spin magnitude, and isotropic in spin directions.
These ``flat'' priors provide a useful starting point in the absence of confident predictions about the shape of these distributions.
However, if we believe that our population model accurately describes the underlying distribution, we can use the population of observations to create a physically informed prior for each of the events.
This physically informed prior for event $j$ is the ``leave-one-out'' posterior-predictive distribution
\begin{equation}
    P(\theta_j | \{d\}_{i \neq j}) = \int d\Lambda \, \pi(\theta | \Lambda) P(\Lambda | \{d\}_{i \neq j}).
\label{eq:louppd}
\end{equation}
Here, $P(\Lambda | \{d\}_{i \neq j})$ is the posterior distribution for the population parameters given all of the observations except for the $j^\text{th}$ event.
We omit the $j^\text{th}$ event to avoid double counting.

The population-informed posterior for the $j^\text{th}$ event is given by
\begin{equation}
    P(\theta_j | \{d\}) = \frac{{\cal L}(d_{j} | \theta_j) P(\theta_j | \{d\}_{i \neq j})}{{\cal Z}(d_j | \{d\}_{i \neq j})}.
\end{equation}
Where the term ${\cal L}$ is the usual single event likelihood, $P$ is the leave-one-out posterior predictive distribution, and ${\cal Z}(d_j | \{d\}_{i \neq j})$ is the Bayesian evidence for the event given all other observations.
A similar method is employed in~\cite{fishbach2019}.
In our version of this analysis, we use the $p_\text{astro}$ weighted population posteriors to compute these updated single event posteriors.

In practice, it is more computationally efficient to compute the leave-one-out posterior predictive distribution from the full posterior predictive distribution which includes all events in the catalog.
To obtain this we rewrite Eq.~\ref{eq:louppd} as
\begin{align}
    P(\theta_j | \{d\}_{i \neq j})
    &= \int d\Lambda \, \pi(\theta_j | \Lambda) P(\Lambda | \{d\}_{i \neq j}) \nonumber \\
    &= \int d\Lambda \, \pi(\theta_j | \Lambda) P(\Lambda | \{d\}) \frac{P(\Lambda | \{d\}_{i \neq j})}{P(\Lambda | \{d\})} \nonumber \\
    &= \frac{{\cal Z}(\{d\})}{{\cal Z}(\{d\}_{i \neq j})}
    \int d\Lambda \, P(\Lambda | \{d\}) \frac{\pi(\theta_j | \Lambda)}{{\cal L}(d_j | \Lambda)} \nonumber \\
    &\appropto \sum_{\{\Lambda_k\}} \frac{\pi(\theta_j | \Lambda_k)}{{\cal L}(d_j | \Lambda_k)}
    ,
    \label{eq:leaveoneout}
\end{align}
where $\Lambda_k$ are the population posterior samples. Going from the second to third lines we use that the combined likelihood is the product of the single event likelihoods.
In the final line the sum is over samples from the population posterior with all events.
Therefore, we obtain the leave-one-out posterior predictive distribution by weighting our posterior samples from the full posterior predictive distribution by the inverse of the single-event marginalized likelihood ${\cal L}(d_j | \Lambda_k)$ as defined in Eq.~\ref{eq:marg_likelihood}.
In order to account for $p_\text{astro}$, we weight the posterior predictive distribution by the inverse of  ${\cal L}_\text{astro}(d_j | \Lambda_k)$, which is the single-event marginalized likelihood weighted by the inverse of $p_\text{astro}(\Lambda)$.

\begin{figure*}
    \begin{subfigure}[t]{\columnwidth}
    \includegraphics[width=0.79\linewidth]{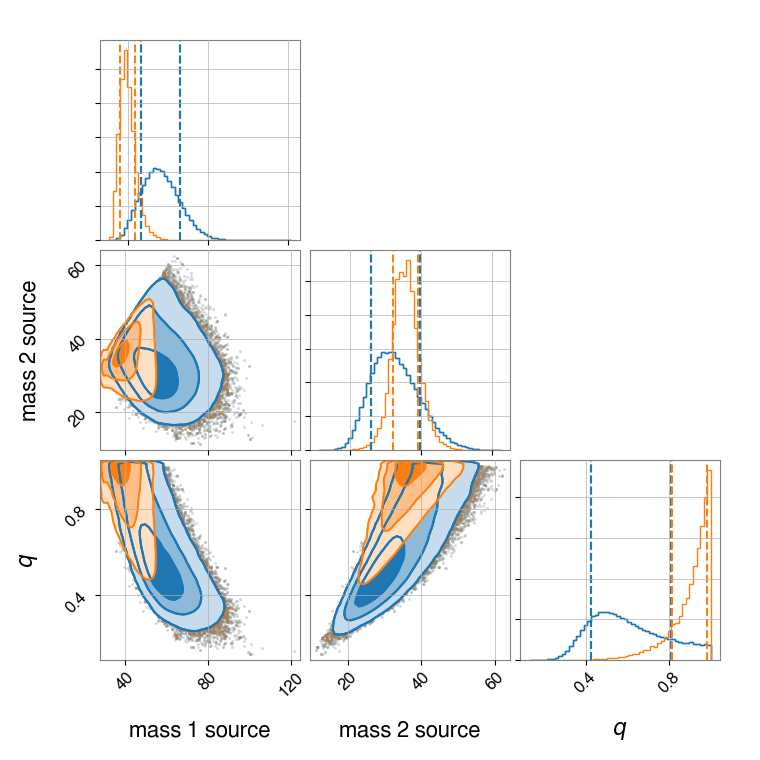} 
    \end{subfigure}
    \begin{subfigure}[t]{\columnwidth}
    \includegraphics[width=\linewidth]{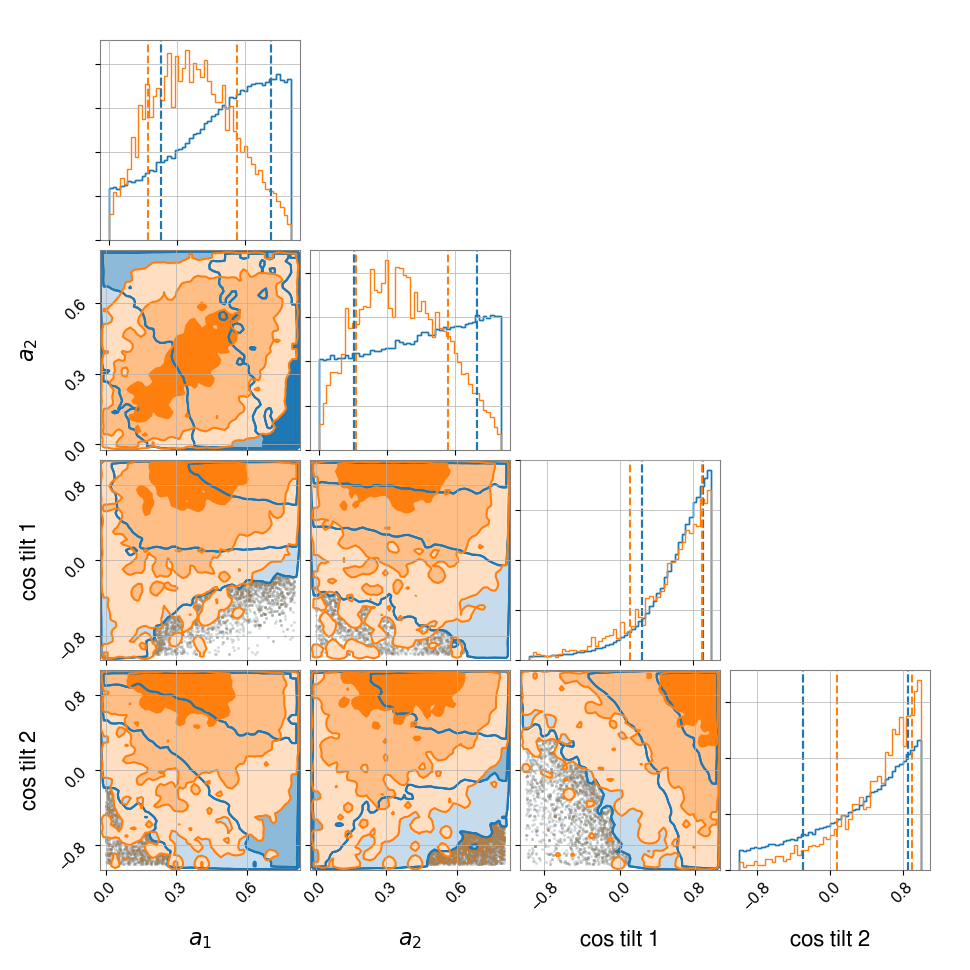}
    \end{subfigure}
    \caption{Posterior distributions for GW170729. The blue contours are calculated using the default priors from \textsc{Bilby} (described in~\ref{overview}), and the orange contours are calculated using a posterior predictive distribution, which uses the other events in GWTC-1 to produce a more astrophysically motivated prior.
    }
    \label{fig:ppd_GW170729}
\end{figure*}

To illustrate this method, we single out GW170729, the most massive event in the GWTC-1 catalog, and a source of speculation about sub-populations of black holes~\cite{NU,Katerina}.
In Fig.~\ref{fig:ppd_GW170729}, we plot the population-weighted posterior distribution for GW170729.
After applying the population-weighting, the posterior for the masses of GW170729 shift toward smaller values.
This result is broadly consistent with the findings from~\citeauthor{fishbach2019}, which showed that the highest-mass events in a gravitational-wave transient catalog generically shifts to smaller values.
We present population-weighted posteriors for the other GWTC-1 events in Appendix~\ref{sec:appendix-astro-prior}.

\subsection{Updating $p_{\text{astro}}$ for GWTC-1}\label{pastro-update}
Using Equation~\ref{eq:calculating_pastro} we calculate updated values of $p_{\text{astro}}$ for each event in GWTC-1. 
The new $p_\text{astro}$ values show how our confidence in the astrophysical origin of each event has changed based on how well it conforms to our population model.

\begin{table}[h!]
\begin{tabular}{c c c c} 
    \hline
    Event & Original $p_\text{astro}$ & Updated $p_\text{astro}$ & Catalog\\ 
    \hline\hline
    GW150914 & 1.00 & 1.00 & GWTC-1\\
    GW151012 & 0.96 & 1.00 & GWTC-1\\
    GW151216 & 0.71 & 1.00 & IAS \\
    GW151226 & 1.00 & 1.00 & GWTC-1\\
    GW170104 & 1.00 & 1.00 & GWTC-1\\
    GW170121 & 1.00 & 1.00 & IAS \\
    GW170202 & 0.68 & 1.00 & IAS \\
    GW170304 & 0.985 & 1.00 & IAS \\
    GW170403 & 0.56 & 1.00 & IAS \\
    GW170425 & 0.77 & 1.00 & IAS \\
    GW170608 & 1.00 & 1.00 & GWTC-1\\
    GW170727 & 0.98 & 1.00 & IAS \\
    GW170729 & 0.52 & 1.00 & GWTC-1\\
    GW170809 & 1.00 & 1.00 & GWTC-1\\
    GW170814 & 1.00 & 1.00 & GWTC-1\\
    GW170817A & 0.86 & 1.00 & IAS \\
    GW170818 & 1.00 & 1.00 & GWTC-1\\
    GW170823 & 1.00 & 1.00 & GWTC-1\\
    \hline
\end{tabular}
\caption{Updated $p_\text{astro}$ values for the binary black hole events in the GWTC-1 and IAS catalogs. The updated values are the median values for all hyper-posterior samples. Note that the distribution of $p_\text{astro}$ for GW170817A includes a long tail, extending to values near zero.}
\label{tab:updated_pastros}
\end{table}

We present the updated the values of $p_\text{astro}$ in Tab.~\ref{tab:updated_pastros}.
The updated values of $p_\text{astro}$ are calculated by averaging over different values of $\Lambda$.
Following the update, the $p_\text{astro}$ values for all events are now $1.00$.
We interpret this to mean that the population model applied here is a better description of reality than the naive model used for the initial detection statements in~\cite{gwtc1}.
We also calculate the standard deviation in $p_\text{astro}$ for different values of $\Lambda$ and determine that it is small: less than $0.001$; except for GW170817A.
For this event, the 90\% credible interval on $p_\text{astro}$ extends to 0.02.

\section{Discussion}\label{discussion}
As the gravitational-wave transient catalog grows, it will include an increasing number of marginal events, some of which are bound to be terrestrial in origin. 
The astrophysical parameters of these terrestrial events tell us about the nature of interferometer noise, but not about astrophysics.
Therefore, it is important to include information about the origin of each event when performing population analyses. 
We describe a formalism, which allows us to weight marginal events in gravitational-wave catalogs, thereby avoiding mistaken inferences drawn from terrestrial events.
We demonstrate our formalism on events from GWTC-1 and the IAS catalog, and obtain qualitatively similar results to previous work \cite{ratepopO2}; the small differences we do observe are interesting.

Our work highlights a number of issues worthy of future exploration.
First, our study calls attention to the need for a pipeline-independent method for calculating $p_\text{astro}$.
At present, events in GWTC-1 are assigned three different values of $p_\text{astro}$ (corresponding to three LIGO/Virgo pipelines) while the IAS catalog provides yet another estimate of $p_\text{astro}$.
Following~\cite{bcr2}, we argue that the probability that an event is astrophysical need not depend on the pipeline used to detect it.
The development of a suitable noise model would enable population inference without the need to arbitrarily chose one value of $p_\text{astro}$ over another as we have done here for illustrative purposes.

Second, in~\ref{EffectiveNoise}, we point out that our current formulation assumes that astrophysical signals occur in the presence of idealized Gaussian noise while terrestrial false positives are due to non-Gaussian artifacts.
While we believe this a reasonable approximation, it would be interesting to extend the framework presented here to include non-Gaussian noise for the astrophysical hypothesis.
This is likely a non-trivial task, but it seems such a framework will eventually become necessary as the detection rate climbs, and low-level non-Gaussianity begins to affect population inference.

Third, in~\ref{reweight}, we analyse the GWTC-1 catalog using population informed priors. Our findings show that GW170729 is potentially not an outlier when weighted by astrophysically informed priors. However, it is important to highlight that weighting with a posterior predictive distribution is only as reliable as the underlying population model. A fair question to ask is whether the model applied here is reasonable and complete. For example, it has been suggested that GW170729 may be the result of hierarchical mergers~\cite{Yang2019}.
This is not accounted for in our population model.
This is an avenue that would be interesting to explore. 

Finally, we noted in~\ref{pastro-update} that the formalism here can be used to provide an improved classification of categories within the astrophysical hypothesis: binary black hole, binary neutron star, neutron star black hole binary, etc; see also~\cite{pastro}.
Given the importance of this classification scheme for electromagnetic follow-up, and given how much we are learning about the population properties of compact objects, this too seems like an interesting area for future development.

\section{Acknowledgements}
We thank Maya Fishbach, Tom Callister and Reed Essick for the useful discussions. 
We thank Cody Messick for the helpful comments on our draft manuscript. 
This work is supported by ARC grants FT150100281 and CE170100004.
Computational resources used for this research was provided by the LIGO Lab computing facilities. 
This research has made use of data, software and/or web tools obtained from the Gravitational Wave Open Science Center (https://www.gw-openscience.org), a service of LIGO Laboratory, the LIGO Scientific Collaboration and the Virgo Collaboration. LIGO is funded by the U.S. National Science Foundation. Virgo is funded by the French Centre National de Recherche Scientifique (CNRS), the Italian Istituto Nazionale della Fisica Nucleare (INFN) and the Dutch Nikhef, with contributions by Polish and Hungarian institutes.

\bibliography{marginal}

\begin{appendix}
\section{Recipes for $p_\text{astro}$}\label{pastro}
In this appendix we contrast different ways that $p_\text{astro}$ can be calculated.
In~\cite{pastro} (see their Eq.~17), which is based on previous work in~\cite{fgmc}, $p_\text{astro}$ is define like so:
\begin{align}\label{eq:kapadia}
    p_\text{astro} = \int_0^\infty d\Lambda_B 
    \int_0^\infty d\Lambda_S
    \, p(\Lambda_B, \Lambda_S|\vec{d}) \frac{\Lambda_S f(d)}{\Lambda b(d)+\Lambda_s f(d)} .
\end{align}
Here, $\Lambda_B$ is the true average number of background events in the data and $\Lambda_S$ is the true average number of signal events in the data.
Meanwhile, $f(d)$ and $b(d)$ are respectively the foreground and background distributions of the data (or some ranking statistic, which is a function of the data).
The distribution $b(d)$ is typically calculated using bootstrap methods such as time-slides, while $f(d)$ can be calculated with injections.
Eq.~\ref{eq:kapadia} is an example of a recipe for calculating $p_\text{astro}$ as defined in Eq.~\ref{eq:pastro}.
An alternative recipe for estimating $p_\text{astro}$, which avoids bootstrap methods is proposed in~\cite{bcr2,BCR1}.
In~\cite{bcr2}, $p_\text{astro}$ is calculated using a conservative noise model in which non-Gaussian noise is modeled as uncorrelated binary signals in two or more observatories.
While these two methods differ significantly in their underlying assumptions, our formalism can be applied equally well to any recipe.

\section{Additional ambiguous event results for IAS events}\label{sec:appendix-singleevent}
\begin{figure*}
    \begin{subfigure}{\columnwidth}
    \includegraphics[width=\linewidth]{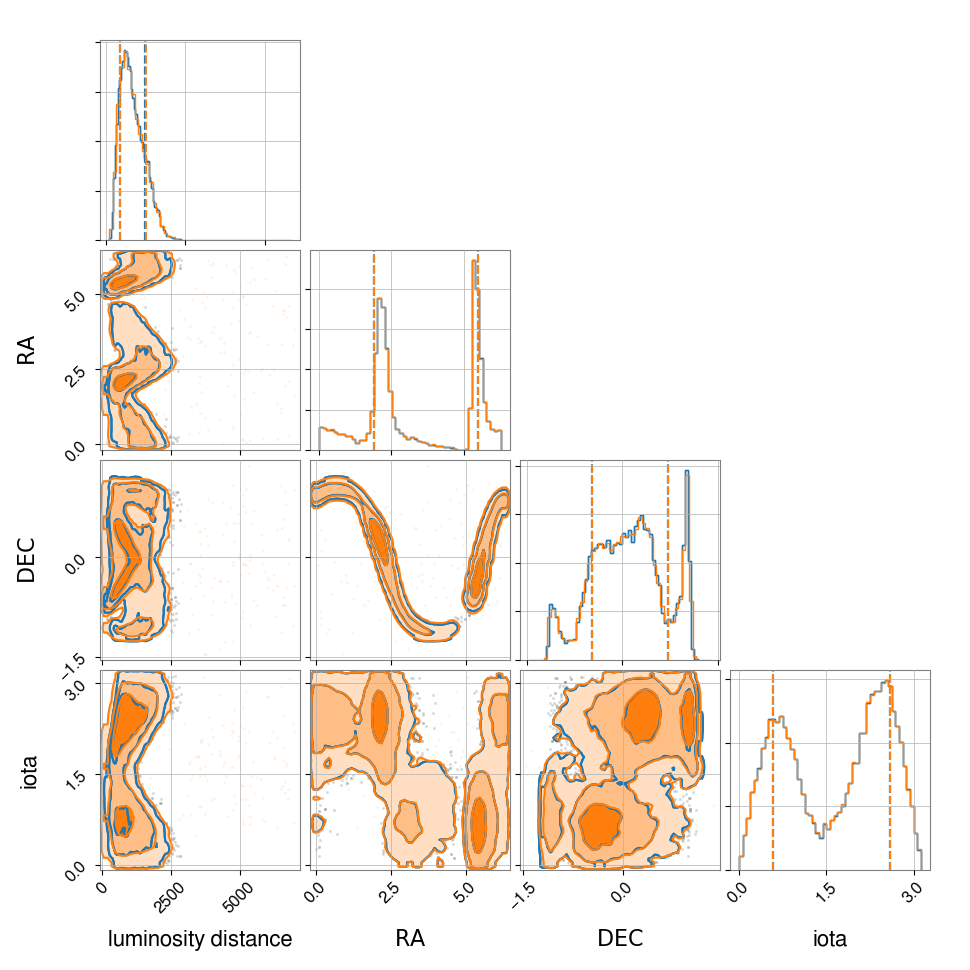} 
    \end{subfigure}
    \begin{subfigure}{\columnwidth}
    \includegraphics[width=\linewidth]{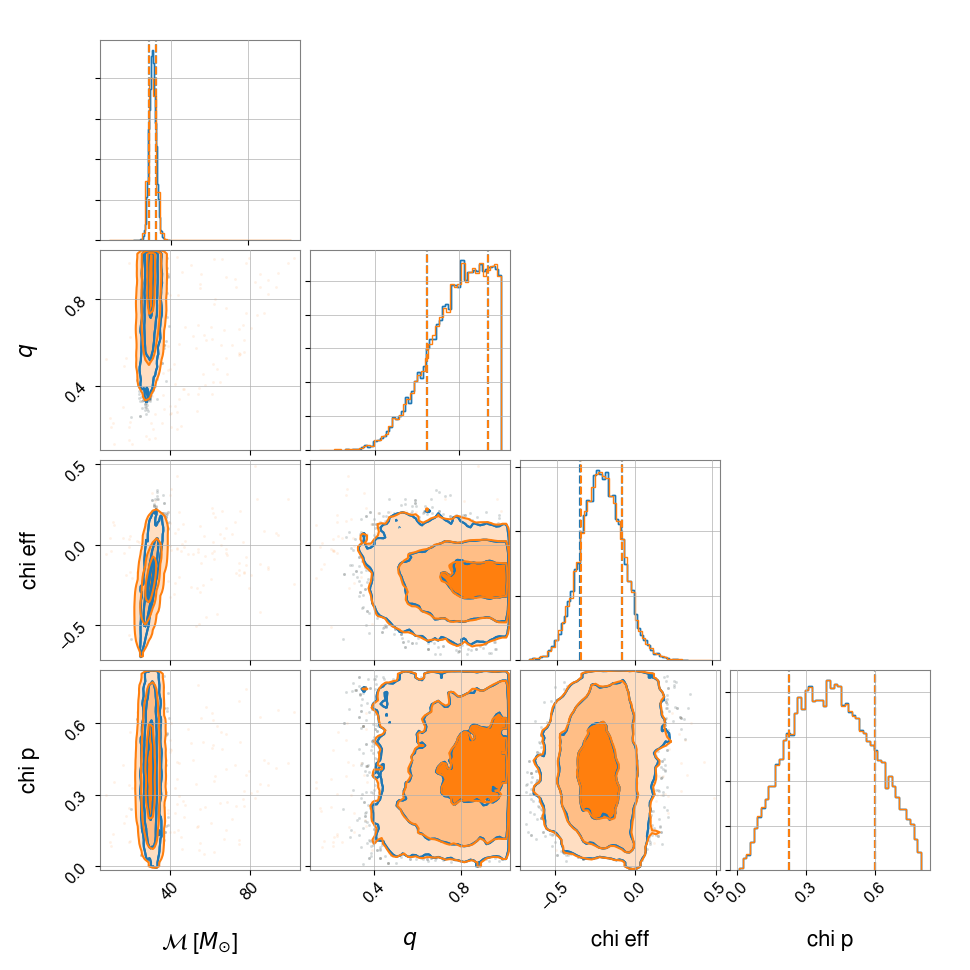}
    \end{subfigure}
    \caption{Posterior distribution of extrinsic (left) and intrinsic (right) parameters for IAS event GW170121 with $p_{\text{astro}}=1.00$.
    Blue contours do not take into account $p_\text{astro}$ while orange contours represent the $p_{\text{astro}}$ weighted posterior distributions.}
    \label{fig:GW170121}
\end{figure*}

\begin{figure*}
    \begin{subfigure}{\columnwidth}
    \includegraphics[width=\linewidth]{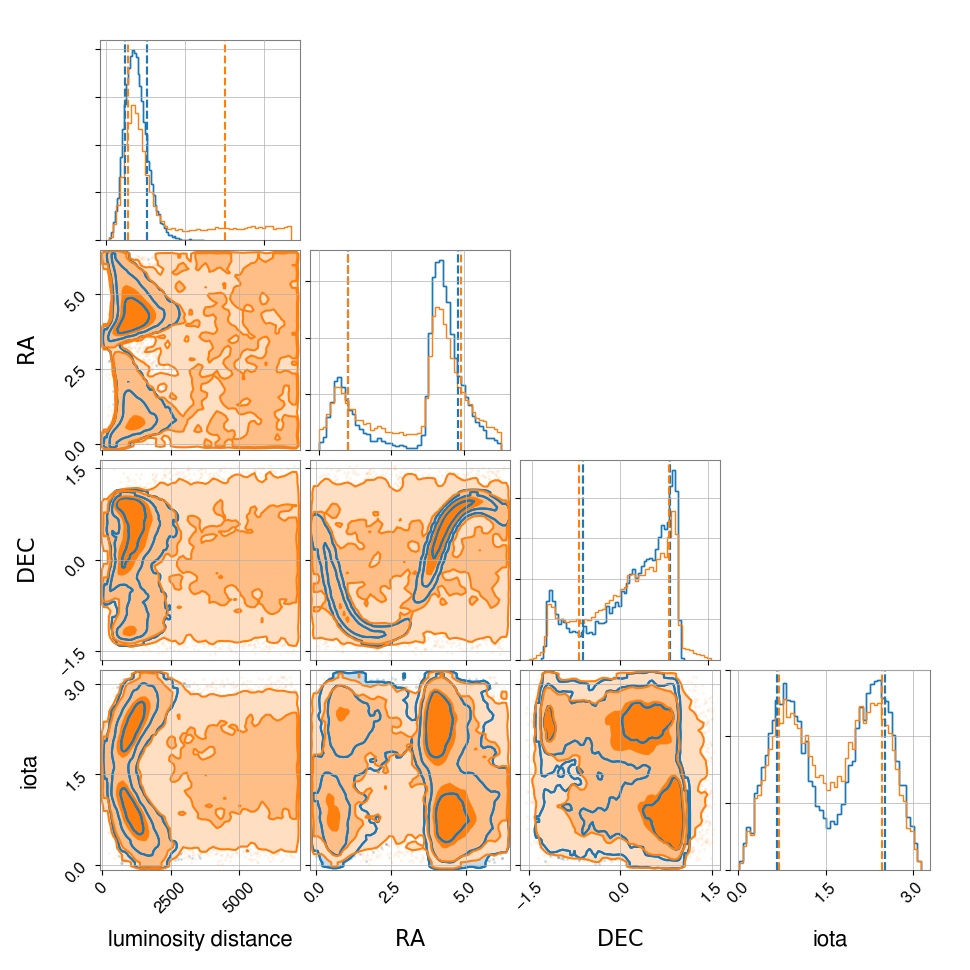} 
    \end{subfigure}
    \begin{subfigure}{\columnwidth}
    \includegraphics[width=\linewidth]{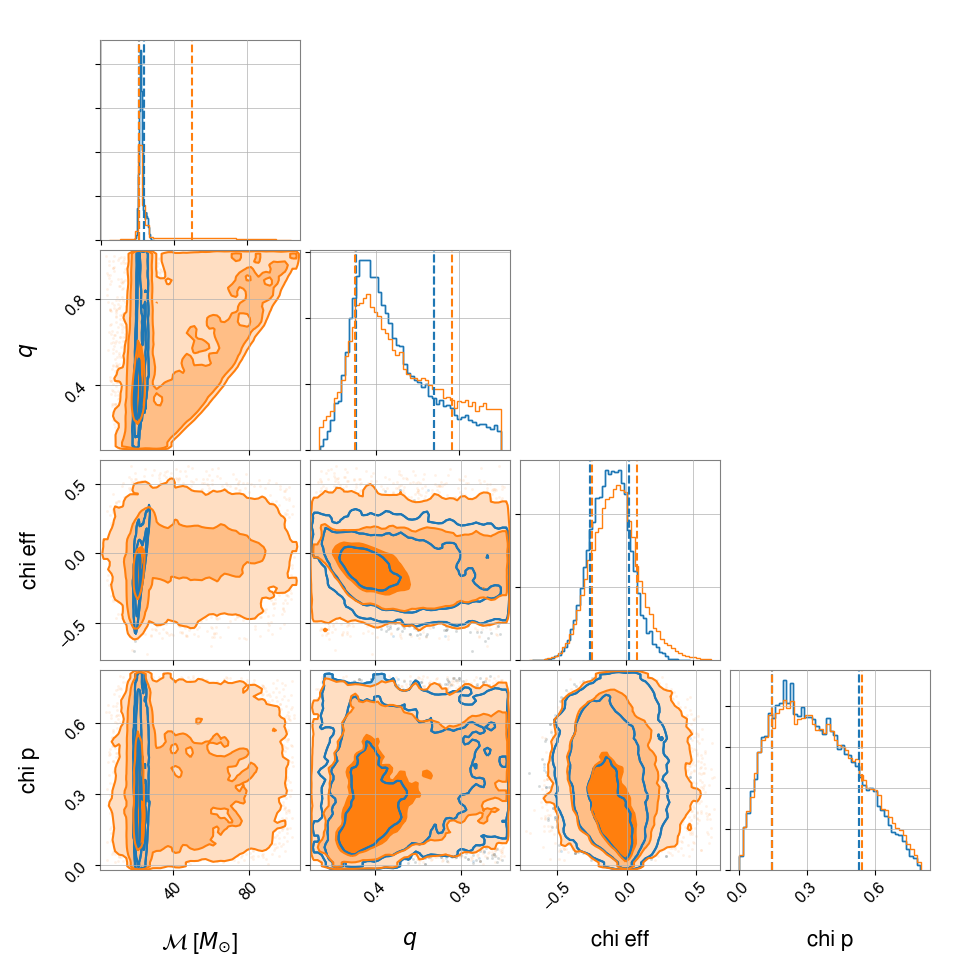}
    \end{subfigure}
    \caption{Posterior distribution of extrinsic (left) and intrinsic (right) parameters for IAS event GW170202 with $p_{\text{astro}}=0.68$.
     Blue contours do not take into account $p_\text{astro}$ while orange contours represent the $p_{\text{astro}}$ weighted posterior distributions.}
    \label{fig:GW170202}
\end{figure*}

\begin{figure*}
    \begin{subfigure}{\columnwidth}
    \includegraphics[width=\linewidth]{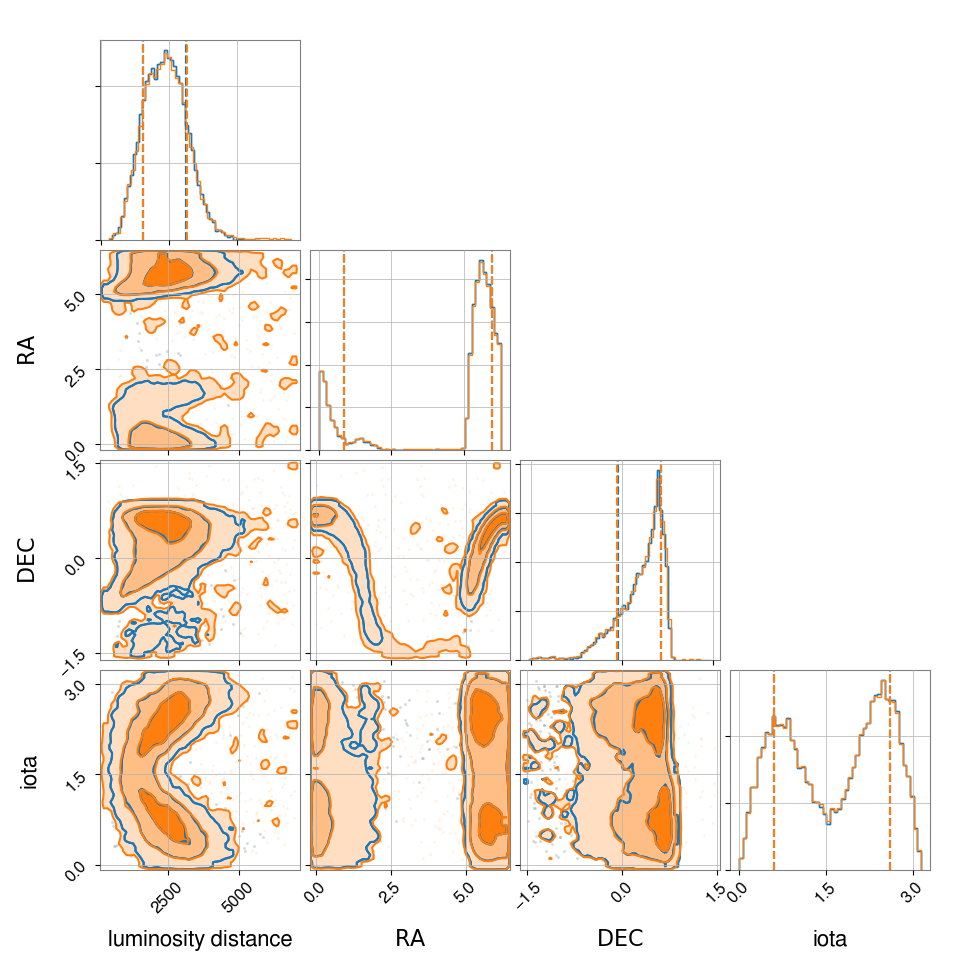} 
    \end{subfigure}
    \begin{subfigure}{\columnwidth}
    \includegraphics[width=\linewidth]{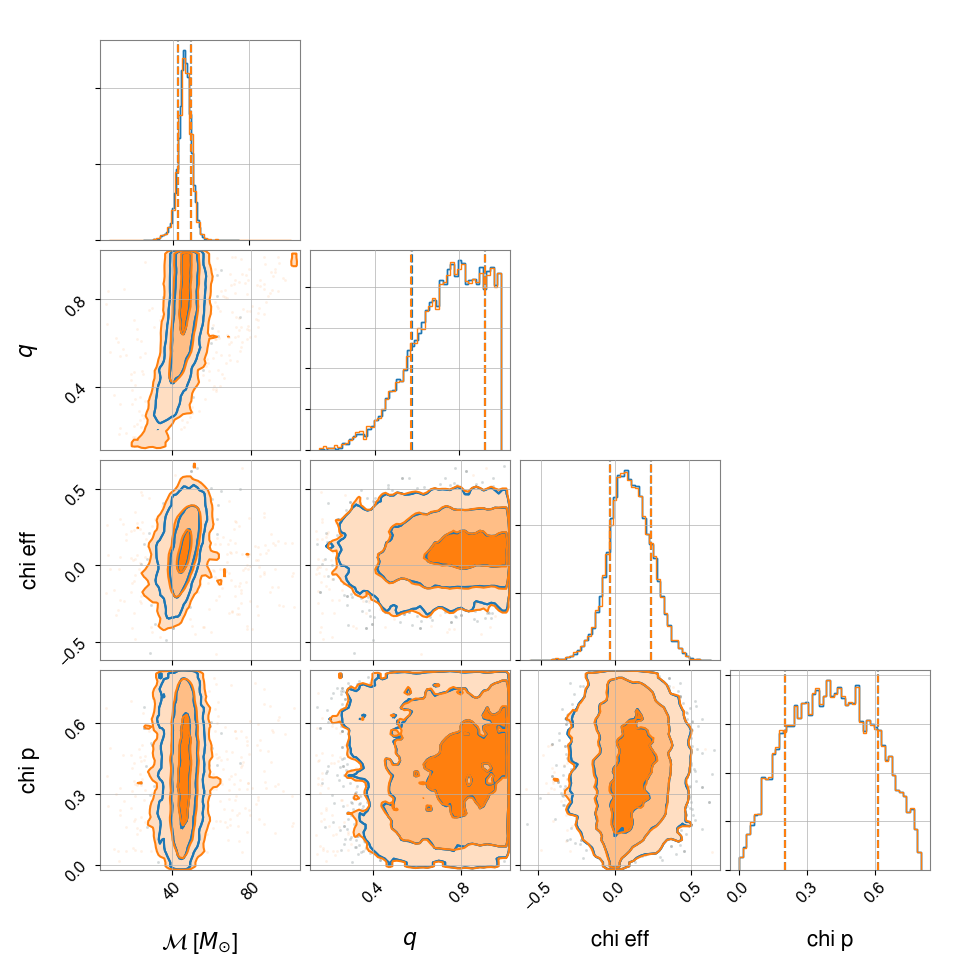}
    \end{subfigure}
    \caption{Posterior distribution of extrinsic (left) and intrinsic (right) parameters for IAS event GW170304 with $p_{\text{astro}}=0.985$.
     Blue contours do not take into account $p_\text{astro}$ while orange contours represent the $p_{\text{astro}}$ weighted posterior distributions.}
    \label{fig:GW170304}
\end{figure*}

\begin{figure*}
    \begin{subfigure}{\columnwidth}
    \includegraphics[width=\linewidth]{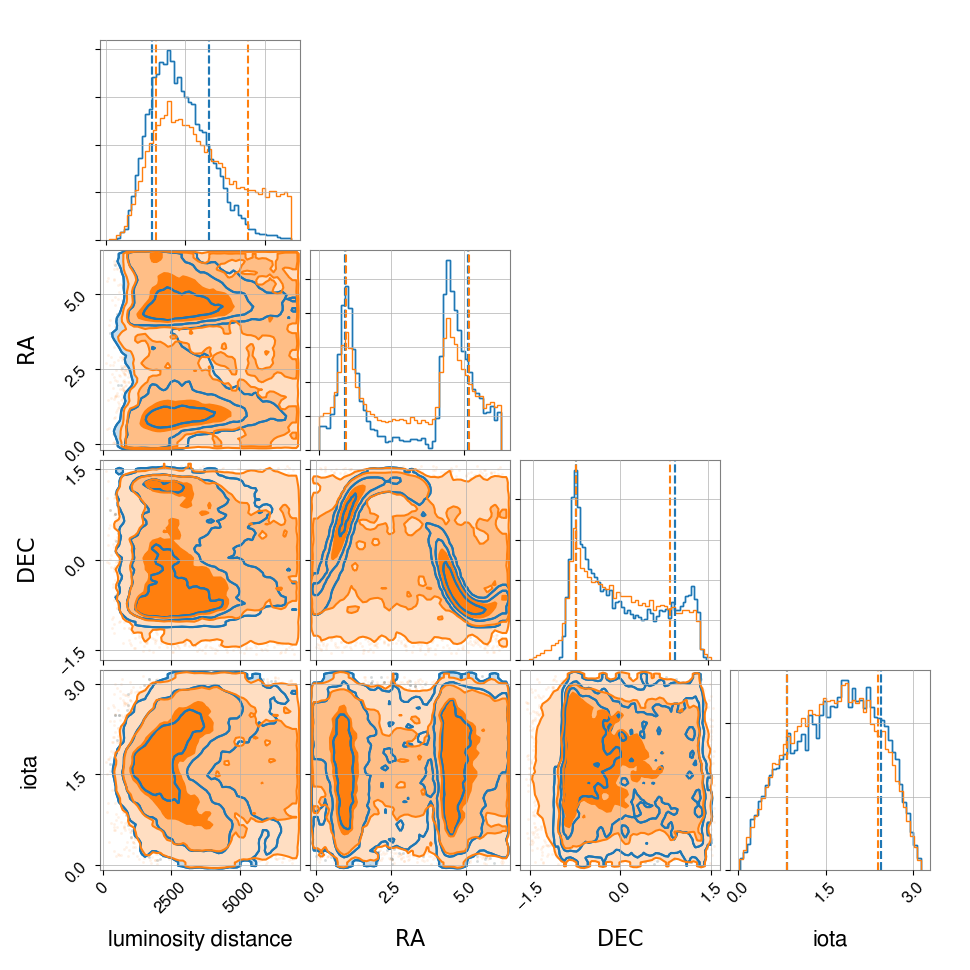} 
    \end{subfigure}
    \begin{subfigure}{\columnwidth}
    \includegraphics[width=\linewidth]{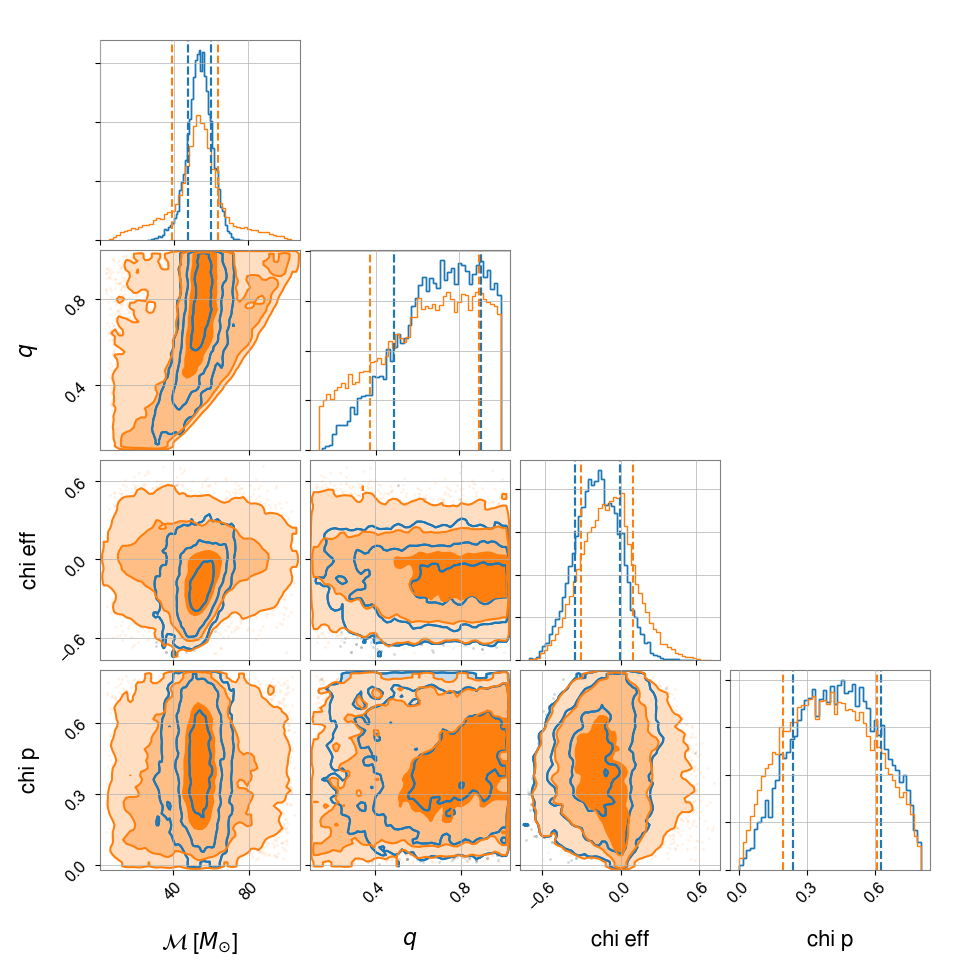}
    \end{subfigure}
    \caption{Posterior distribution of extrinsic (left) and intrinsic (right) parameters for IAS event GW170403 with $p_{\text{astro}}=0.56$.
     Blue contours do not take into account $p_\text{astro}$ while orange contours represent the $p_{\text{astro}}$ weighted posterior distributions.}
    \label{fig:GW170403}
\end{figure*}

\begin{figure*}
    \begin{subfigure}{\columnwidth}
    \includegraphics[width=\linewidth]{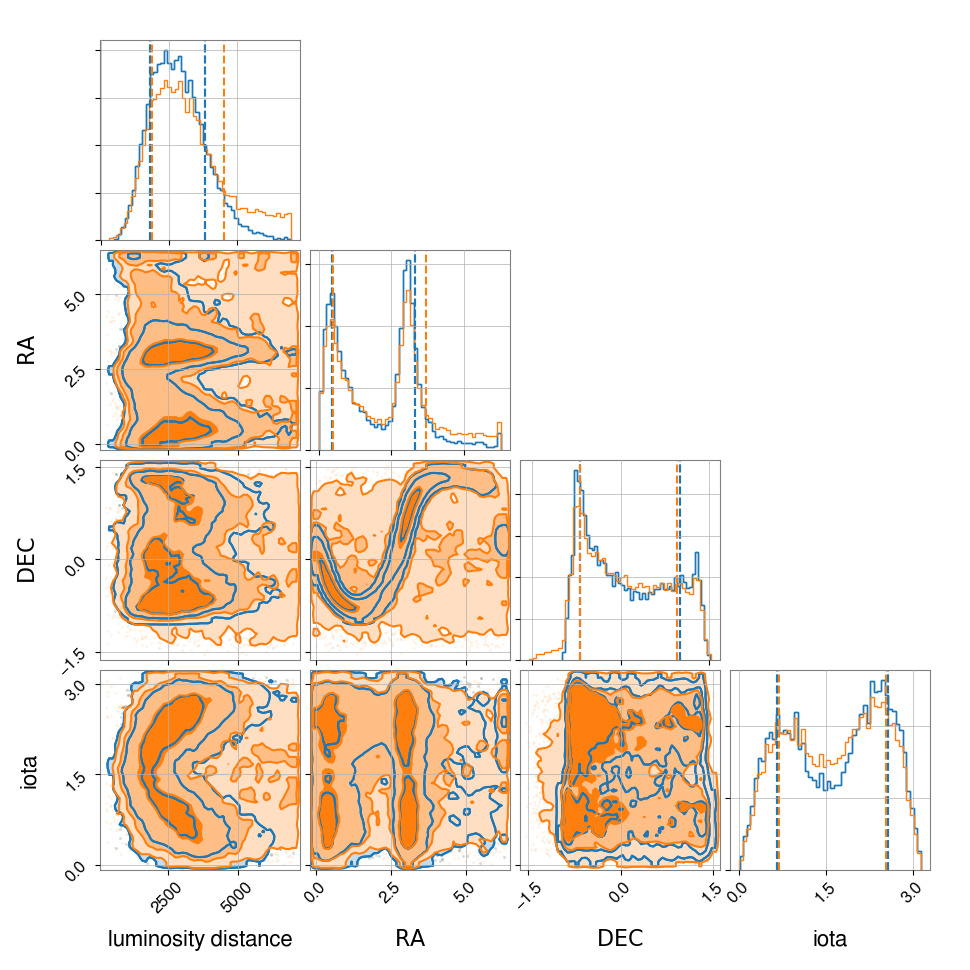} 
    \end{subfigure}
    \begin{subfigure}{\columnwidth}
    \includegraphics[width=\linewidth]{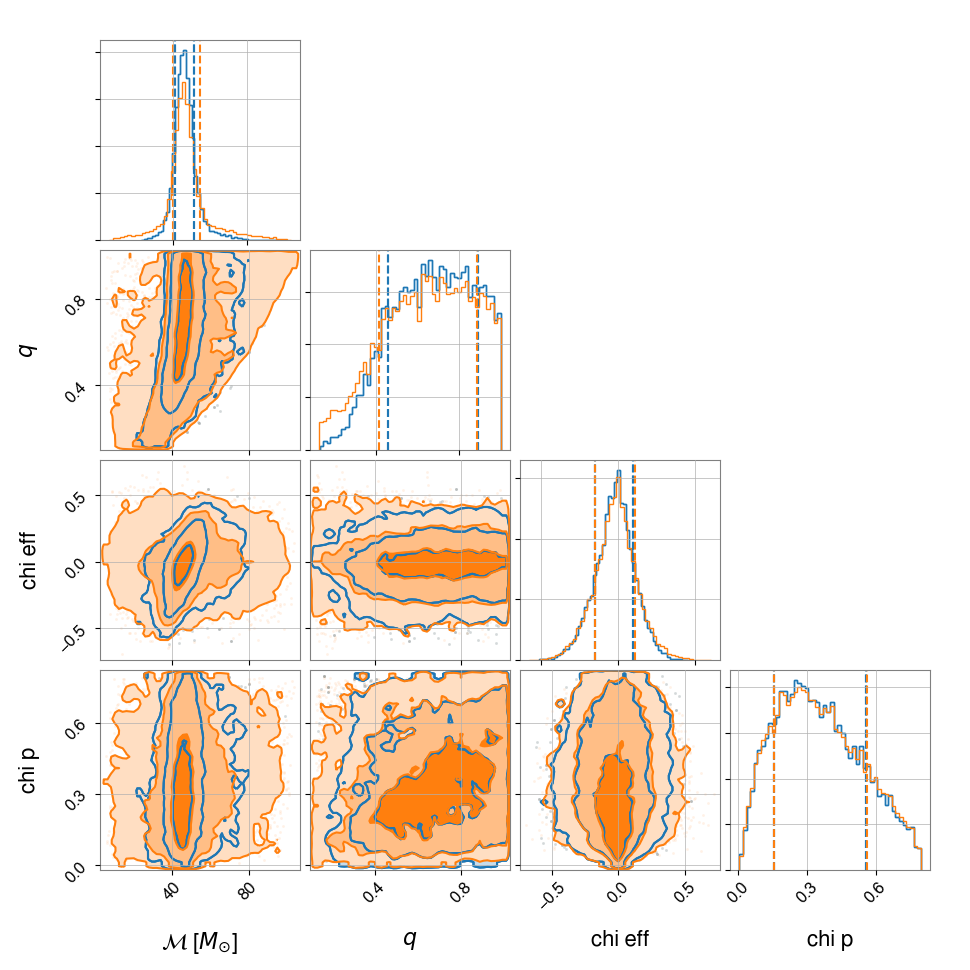}
    \end{subfigure}
    \caption{Posterior distribution of extrinsic (left) and intrinsic (right) parameters for IAS event GW170425 with $p_{\text{astro}}=0.77$.
     Blue contours do not take into account $p_\text{astro}$ while orange contours represent the $p_{\text{astro}}$ weighted posterior distributions.}
    \label{fig:GW170425}
\end{figure*}

\begin{figure*}
    \begin{subfigure}{\columnwidth}
    \includegraphics[width=\linewidth]{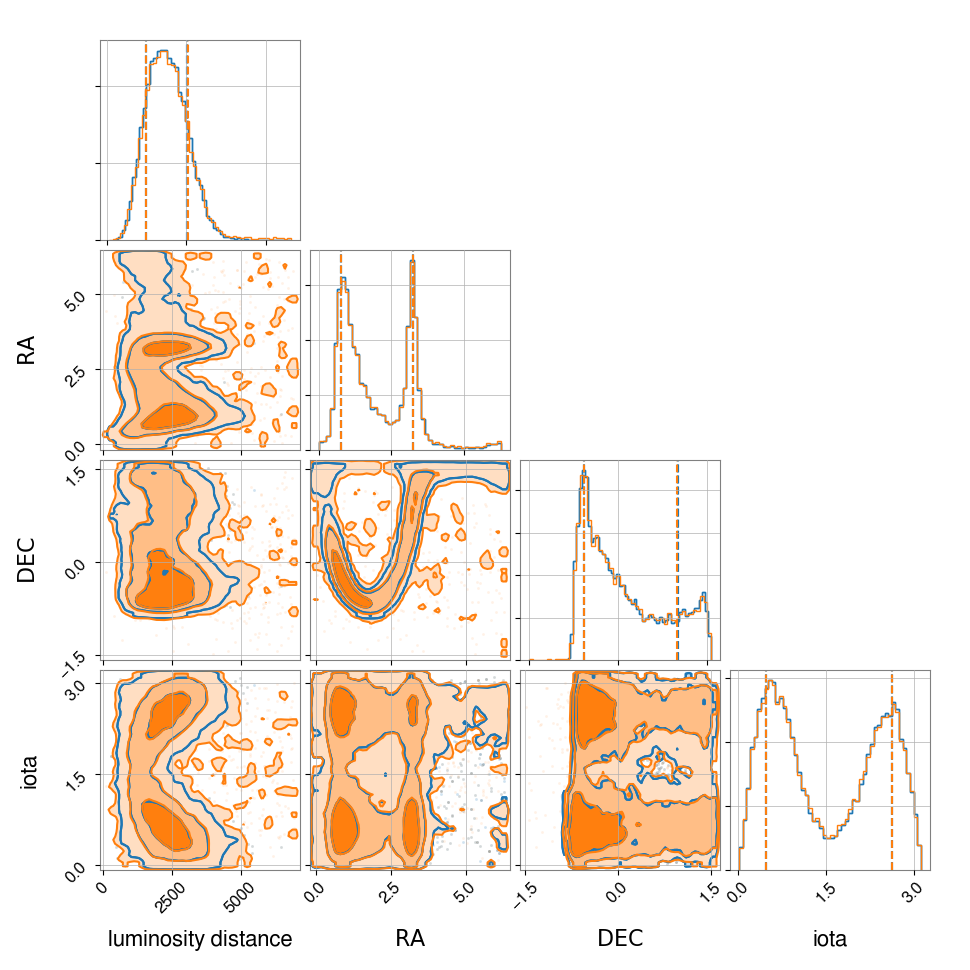} 
    \end{subfigure}
    \begin{subfigure}{\columnwidth}
    \includegraphics[width=\linewidth]{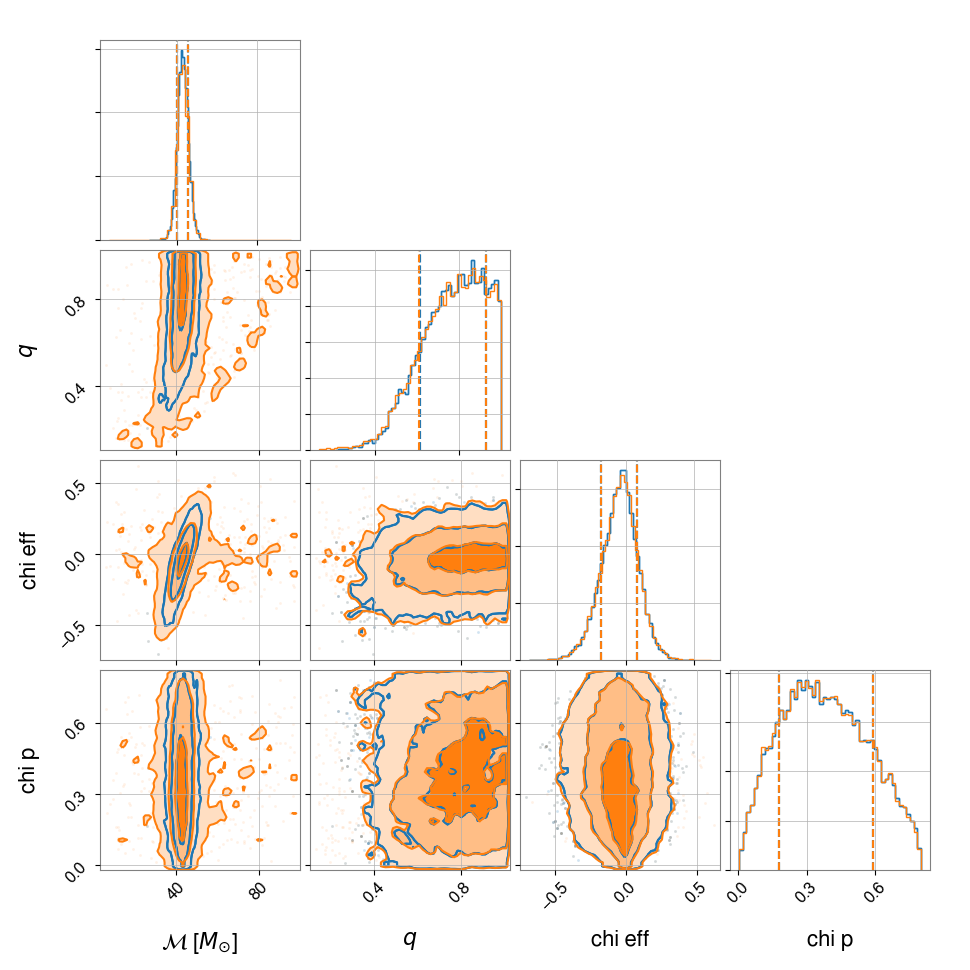}
    \end{subfigure}
    \caption{Posterior distribution of extrinsic (left) and intrinsic (right) parameters for IAS event GW170727 with $p_{\text{astro}}=0.98$.
     Blue contours do not take into account $p_\text{astro}$ while orange contours represent the $p_{\text{astro}}$ weighted posterior distributions.}
    \label{fig:GW170727}
\end{figure*}

\begin{figure*}
    \begin{subfigure}{\columnwidth}
    \includegraphics[width=\linewidth]{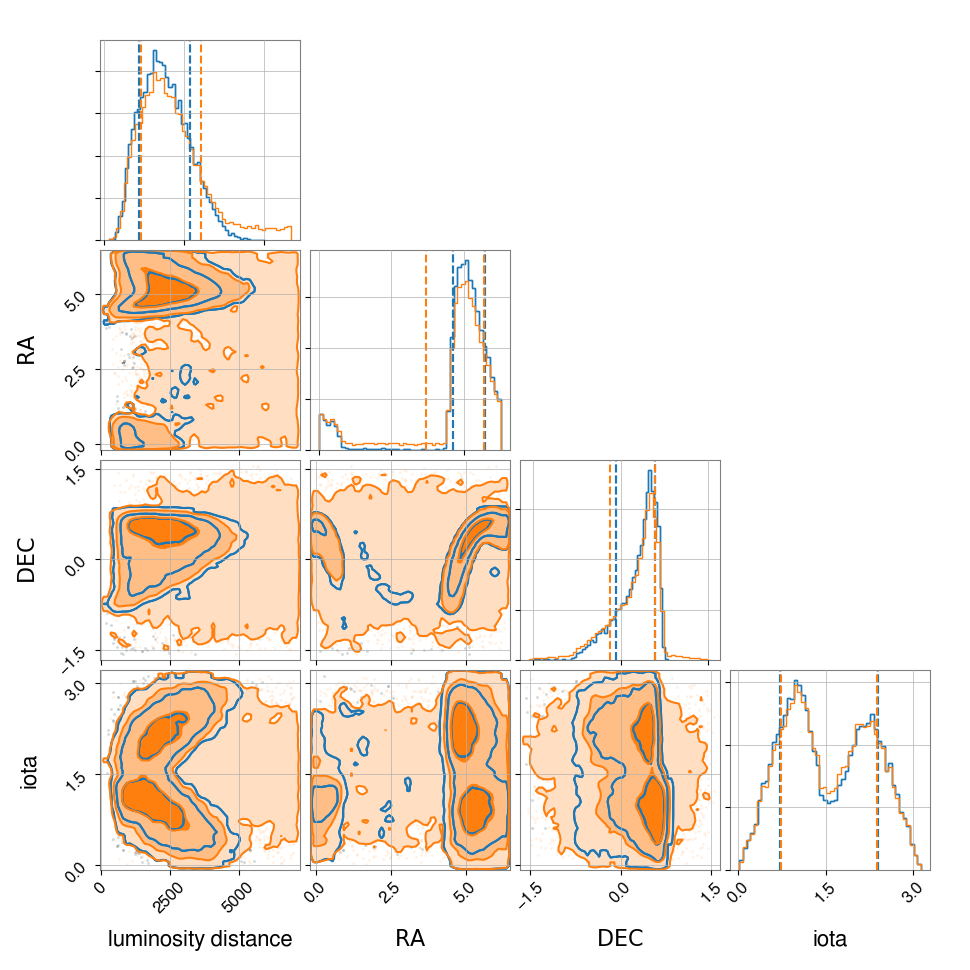} 
    \end{subfigure}
    \begin{subfigure}{\columnwidth}
    \includegraphics[width=\linewidth]{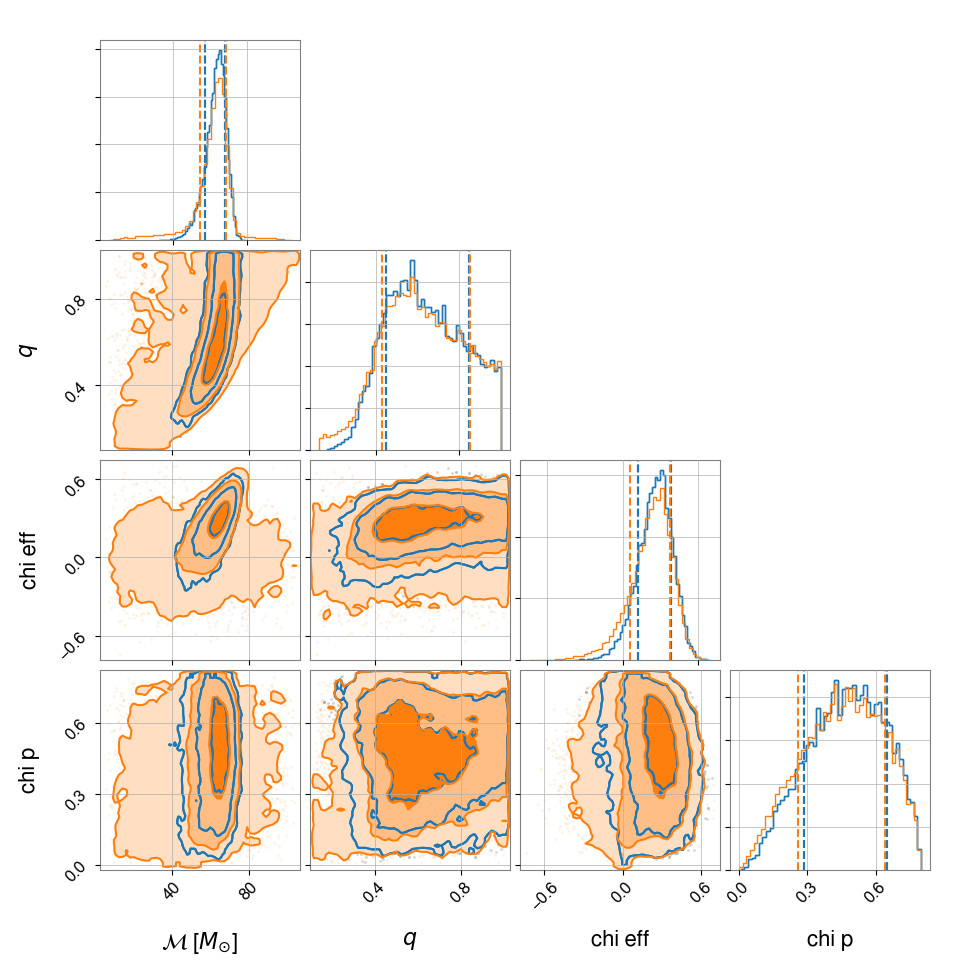}
    \end{subfigure}
    \caption{Posterior distribution of extrinsic (left) and intrinsic (right) parameters for IAS event GW170817A with $p_{\text{astro}}=0.86$.
     Blue contours do not take into account $p_\text{astro}$ while orange contours represent the $p_{\text{astro}}$ weighted posterior distributions.}
    \label{fig:GW170817A}
\end{figure*}

In Figures~\ref{fig:GW170121}-\ref{fig:GW170817A}, we present the $p_\text{astro}$ weighted posteriors for the remaining IAS events \cite{ias3, ias4}.
The blue contours represent the posterior distributions without taking into account $p_\text{astro}$ while the orange contours represent the $p_{\text{astro}}$ weighted posterior distributions. 

\section{Reanalysis of GWTC-1 events with an astrophysically-motivated prior}\label{sec:appendix-astro-prior}
\begin{figure*}
    \begin{subfigure}[t]{\columnwidth}
    \includegraphics[width=0.79\linewidth]{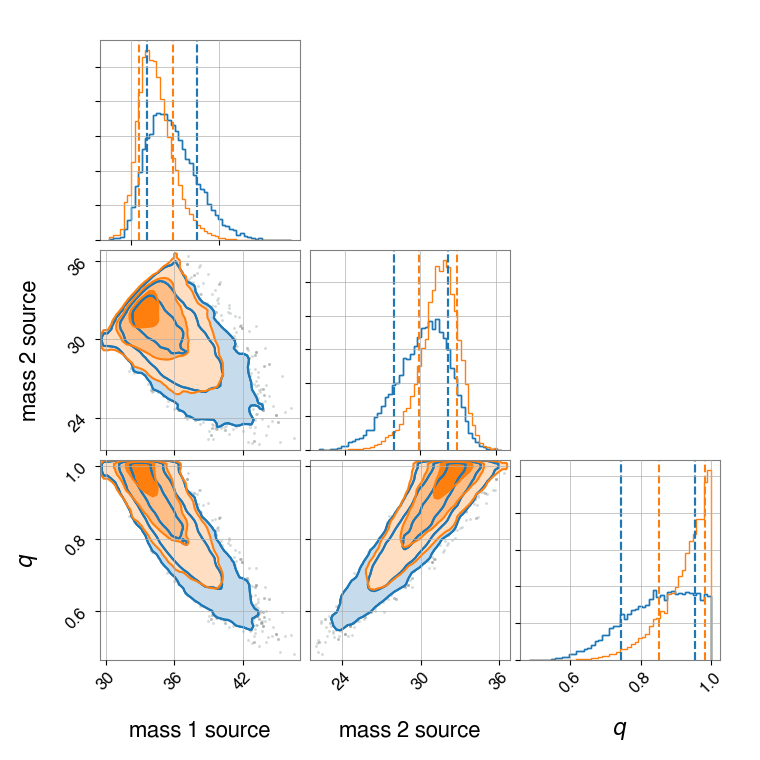} 
    \end{subfigure}
    \begin{subfigure}[t]{\columnwidth}
    \includegraphics[width=\linewidth]{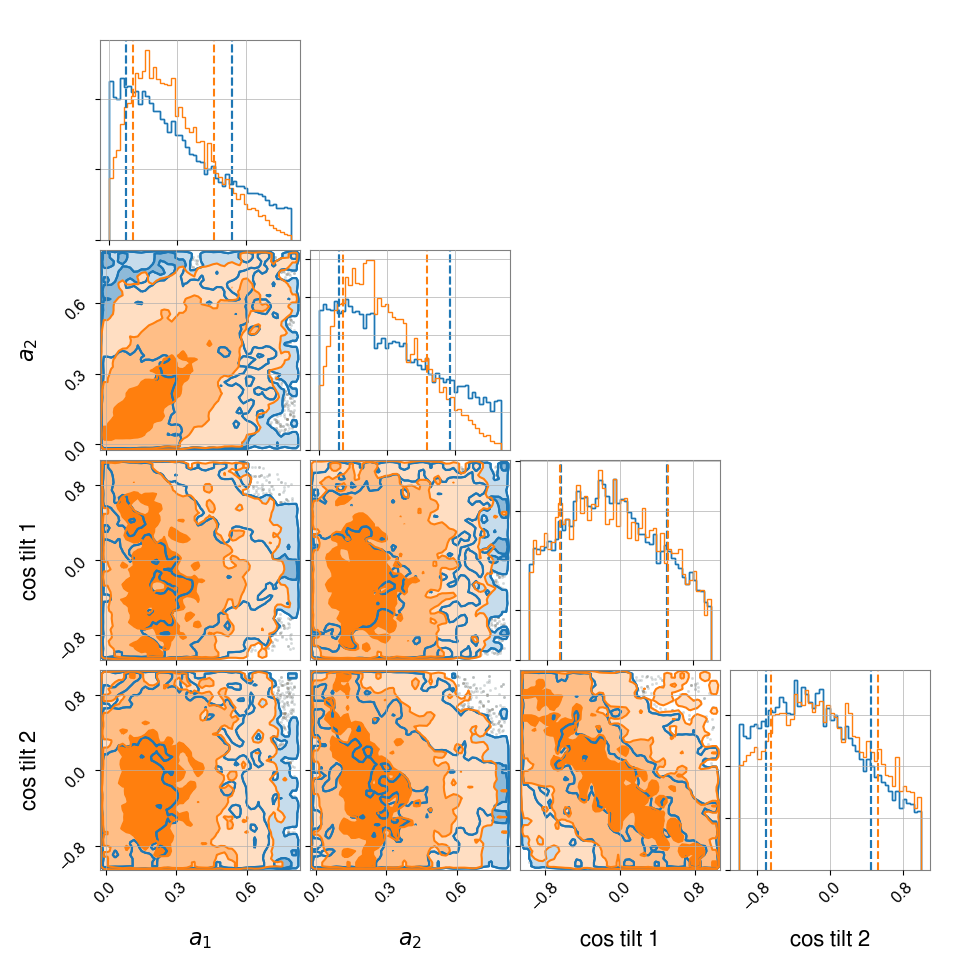}
    \end{subfigure}
    \caption{Posterior distributions for GW150914.
    Blue contours are calculated with uninformative priors while the orange contours are calculated using a posterior predictive distribution.}
    \label{fig:ppd_GW150914}
\end{figure*}

\begin{figure*}
    \begin{subfigure}[t]{\columnwidth}
    \includegraphics[width=0.79\linewidth]{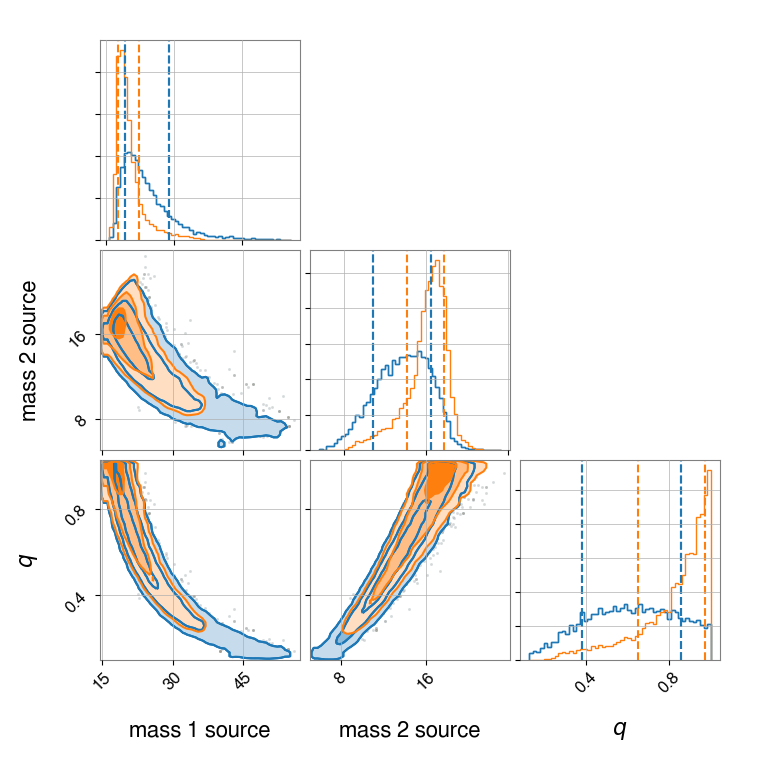} 
    \end{subfigure}
    \begin{subfigure}[t]{\columnwidth}
    \includegraphics[width=\linewidth]{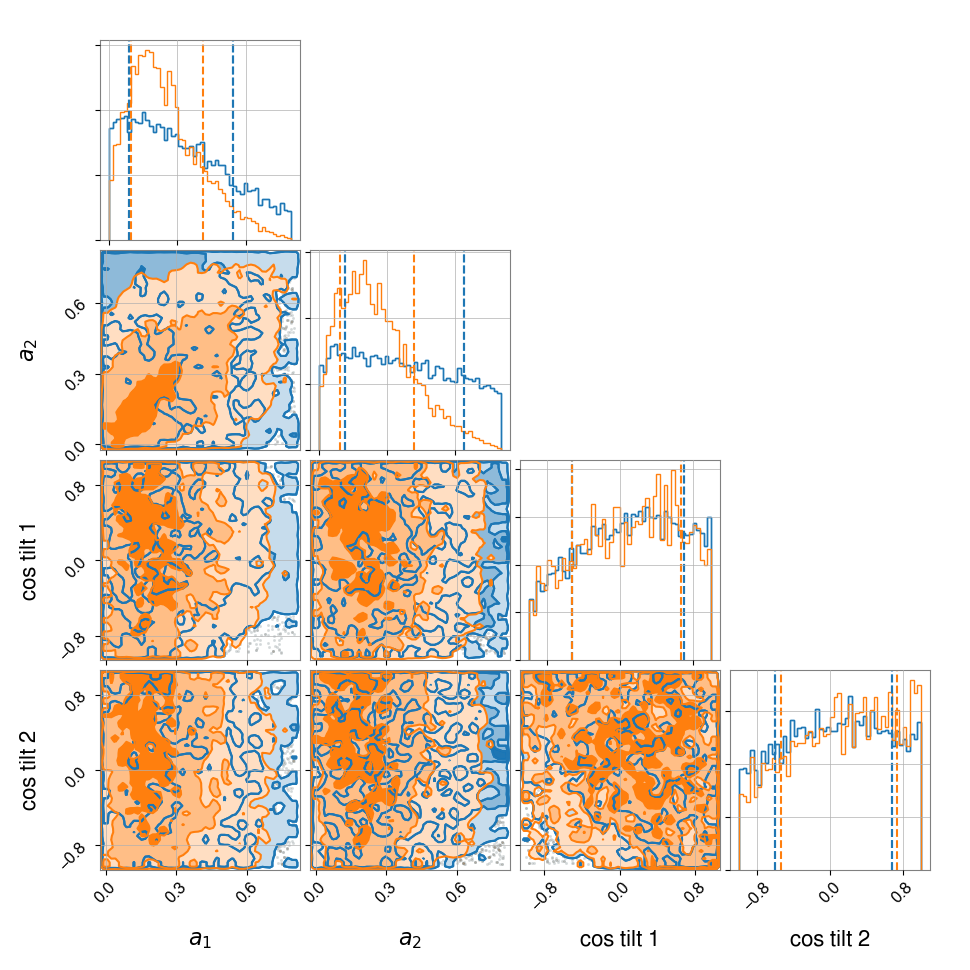}
    \end{subfigure}
    \caption{Posterior distributions for GW151012. Blue contours are calculated with uninformative priors while the orange contours are calculated using a posterior predictive distribution.}
    \label{fig:ppd_GW151012}
\end{figure*}

\begin{figure*}
    \begin{subfigure}[t]{\columnwidth}
    \includegraphics[width=0.79\linewidth]{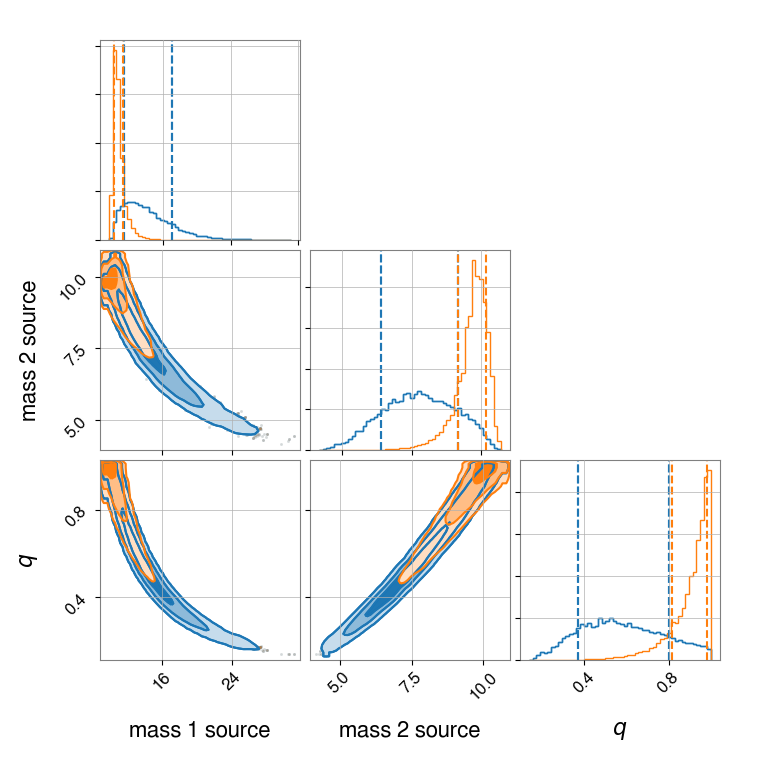} 
    \end{subfigure}
    \begin{subfigure}[t]{\columnwidth}
    \includegraphics[width=\linewidth]{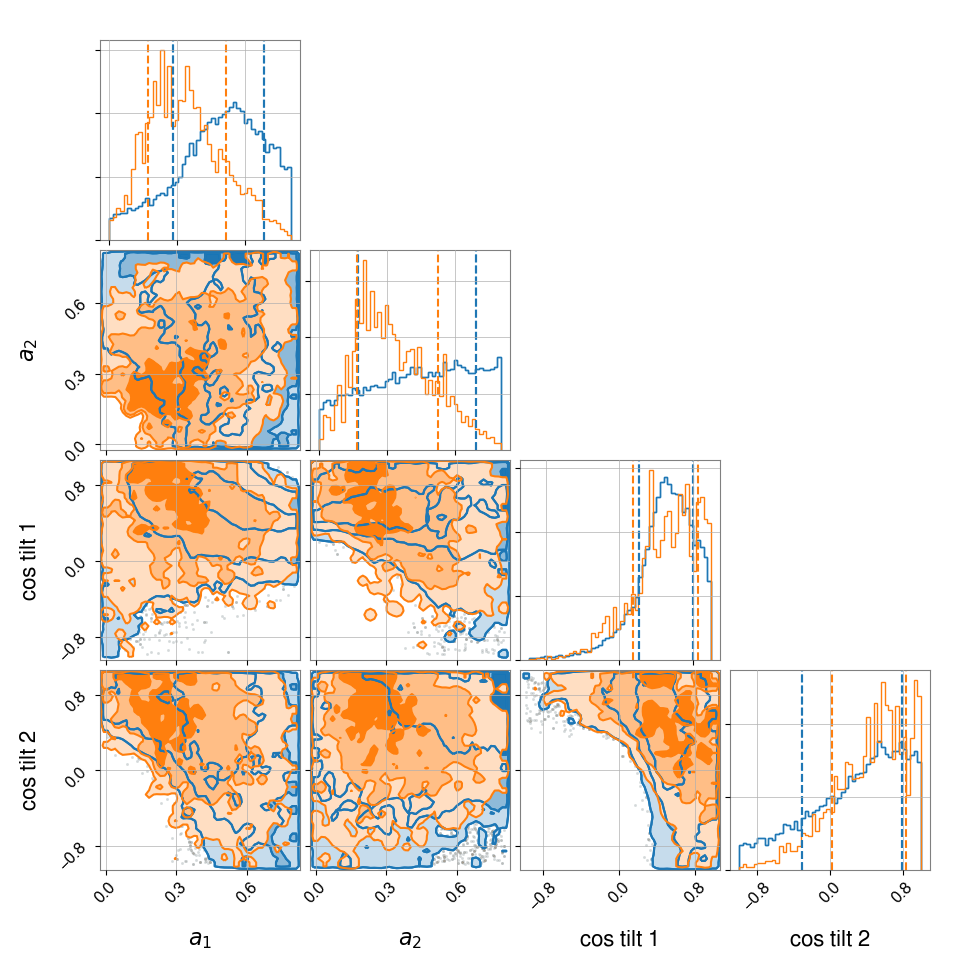}
    \end{subfigure}
    \caption{Posterior distributions for GW151226. Blue contours are calculated with uninformative priors while the orange contours are calculated using a posterior predictive distribution.}
    \label{fig:ppd_GW151226}
\end{figure*}

\begin{figure*}
    \begin{subfigure}[t]{\columnwidth}
    \includegraphics[width=0.79\linewidth]{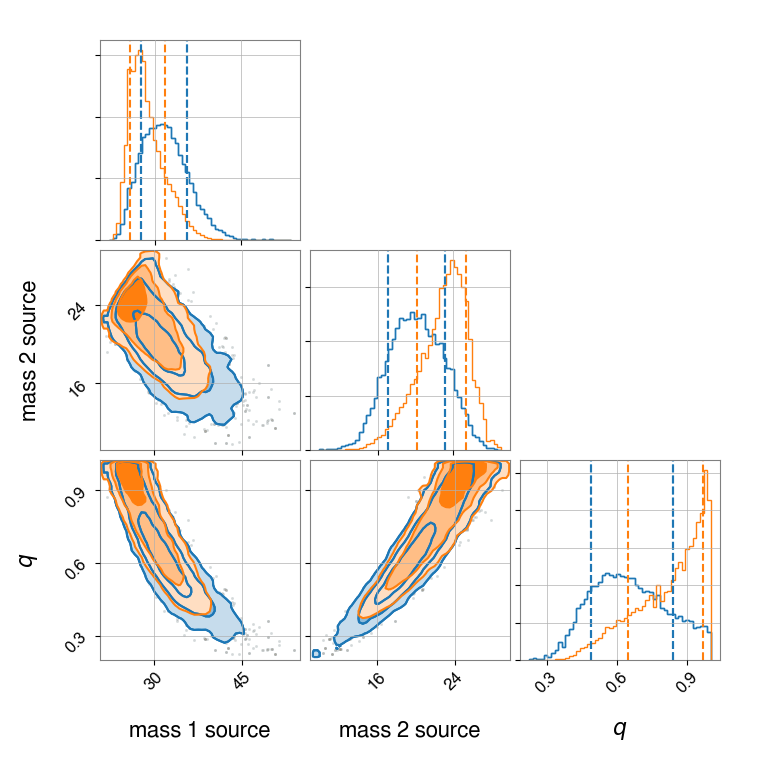} 
    \end{subfigure}
    \begin{subfigure}[t]{\columnwidth}
    \includegraphics[width=\linewidth]{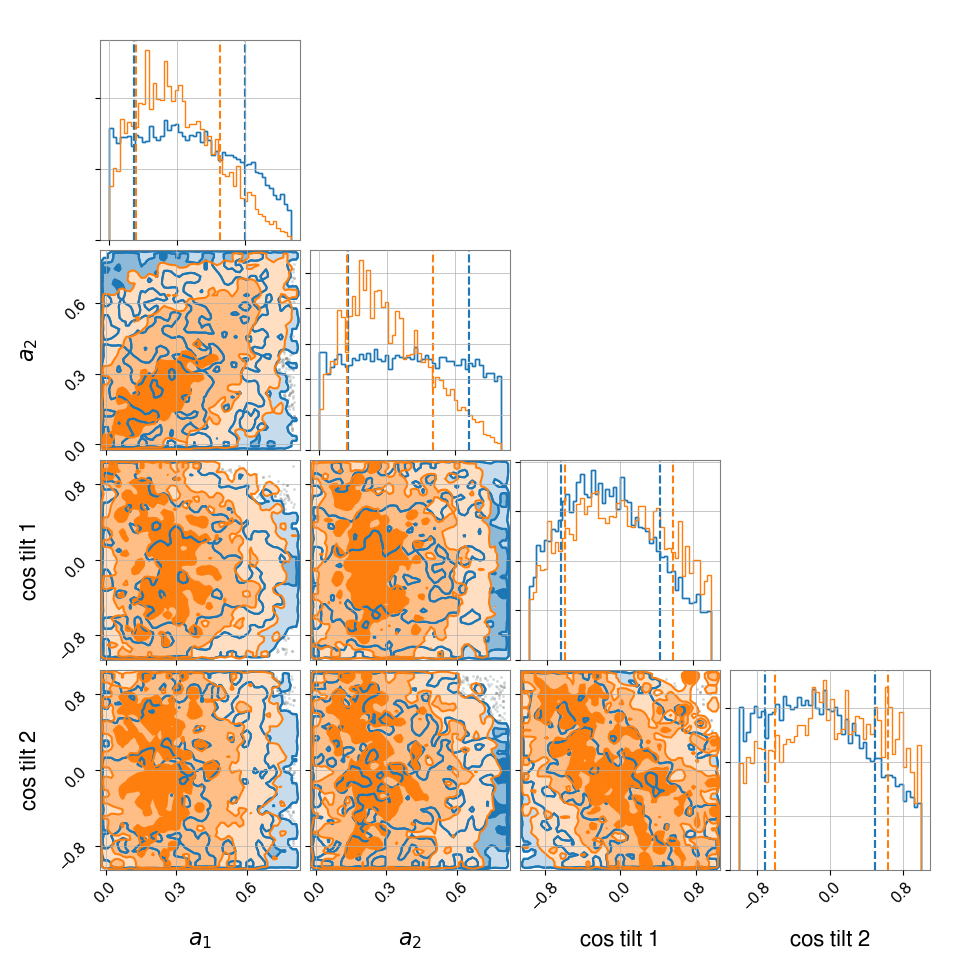}
    \end{subfigure}
    \caption{Posterior distributions for GW170104. Blue contours are calculated with uninformative priors while the orange contours are calculated using a posterior predictive distribution.}
    \label{fig:ppd_GW170104}
\end{figure*}

\begin{figure*}
    \begin{subfigure}[t]{\columnwidth}
    \includegraphics[width=0.79\linewidth]{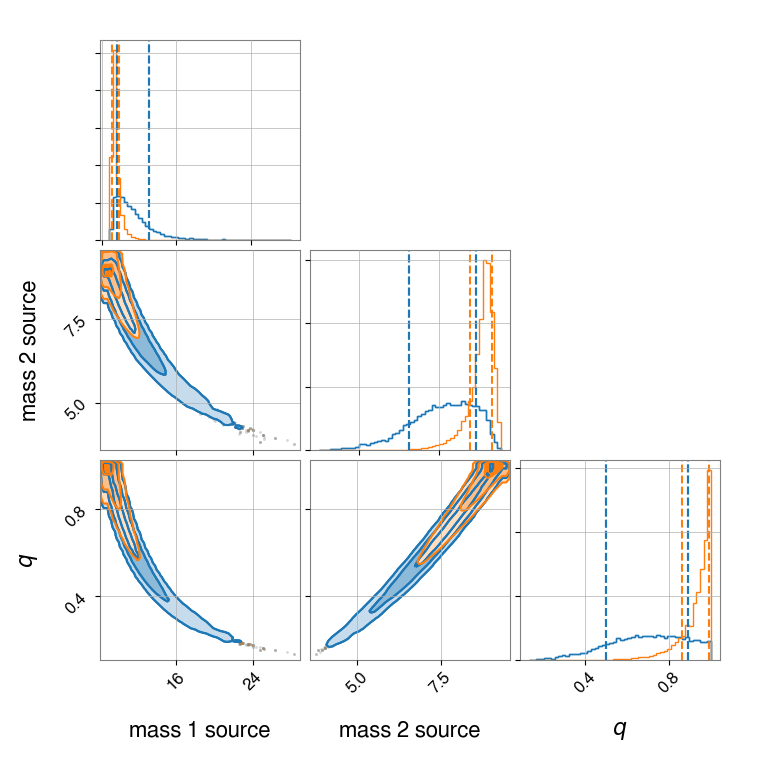} 
    \end{subfigure}
    \begin{subfigure}[t]{\columnwidth}
    \includegraphics[width=\linewidth]{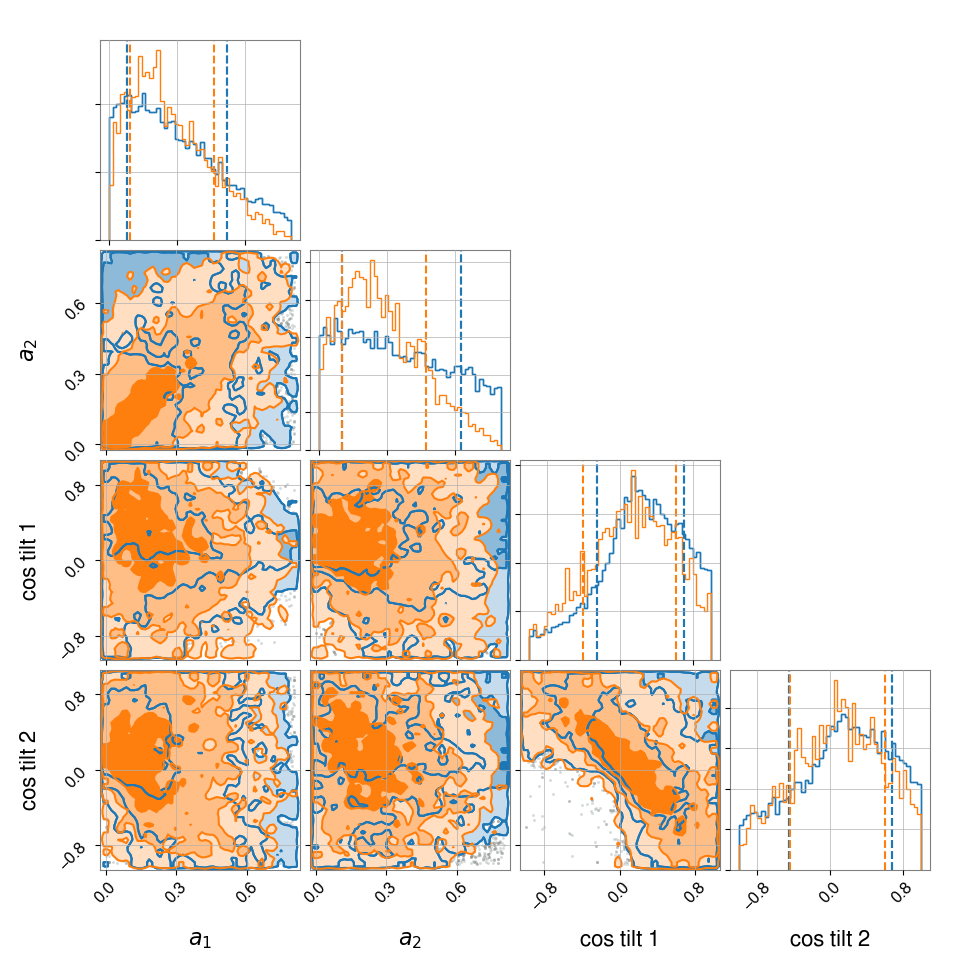}
    \end{subfigure}
    \caption{Posterior distributions for GW170608. Blue contours are calculated with uninformative priors while the orange contours are calculated using a posterior predictive distribution.}
    \label{fig:ppd_GW170608}
\end{figure*}

\begin{figure*}
    \begin{subfigure}[t]{\columnwidth}
    \includegraphics[width=0.79\linewidth]{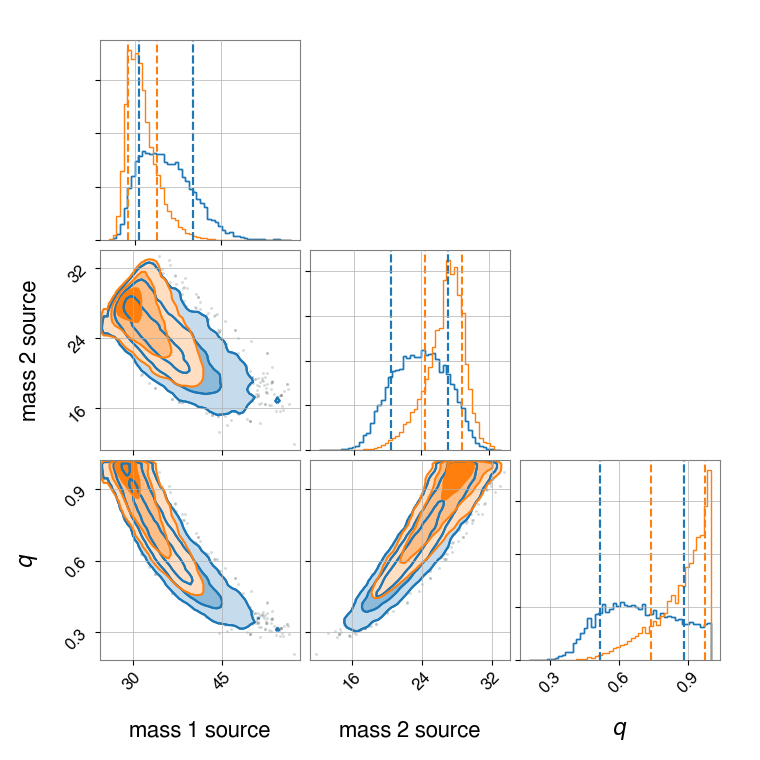} 
    \end{subfigure}
    \begin{subfigure}[t]{\columnwidth}
    \includegraphics[width=\linewidth]{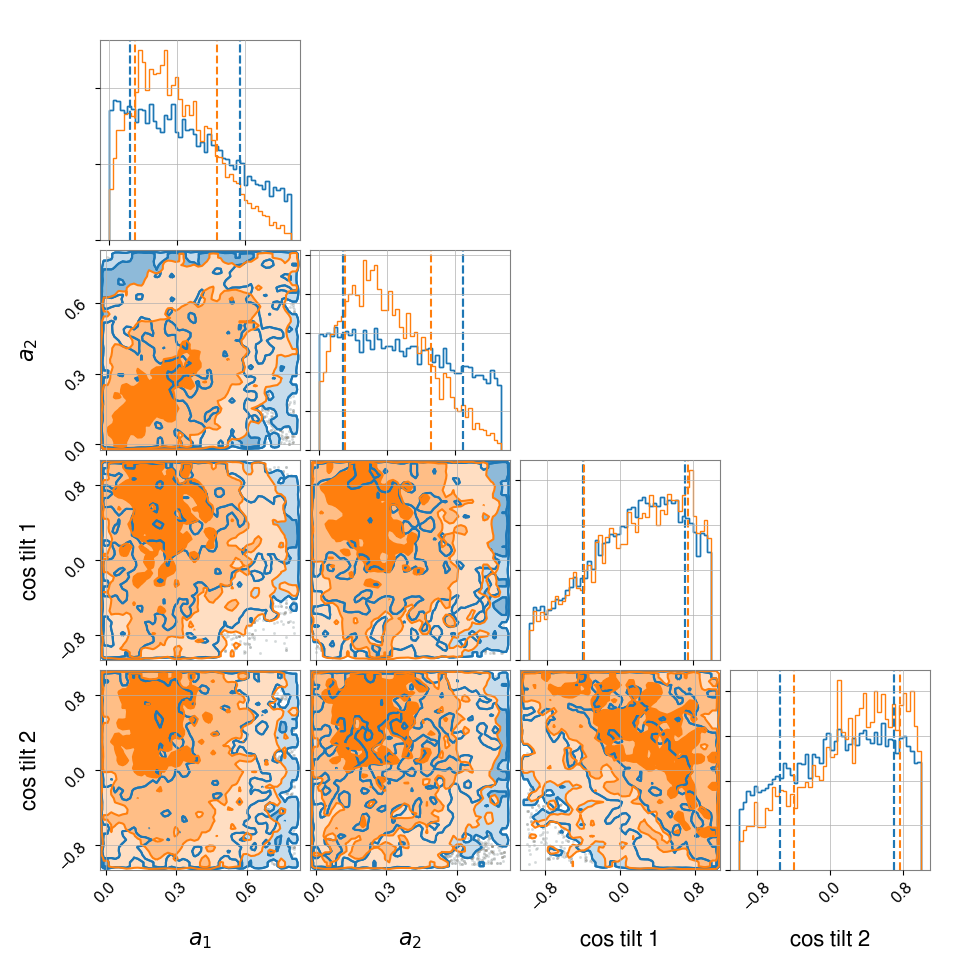}
    \end{subfigure}
    \caption{Posterior distributions for GW170809. Blue contours are calculated with uninformative priors while the orange contours are calculated using a posterior predictive distribution.}
    \label{fig:ppd_GW170809}
\end{figure*}

\begin{figure*}
    \begin{subfigure}[t]{\columnwidth}
    \includegraphics[width=0.79\linewidth]{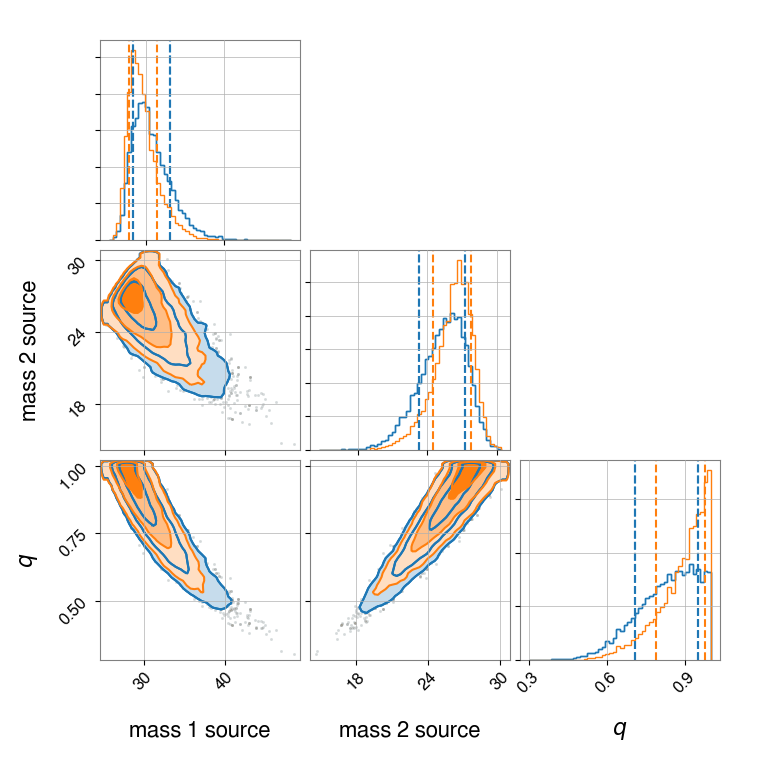} 
    \end{subfigure}
    \begin{subfigure}[t]{\columnwidth}
    \includegraphics[width=\linewidth]{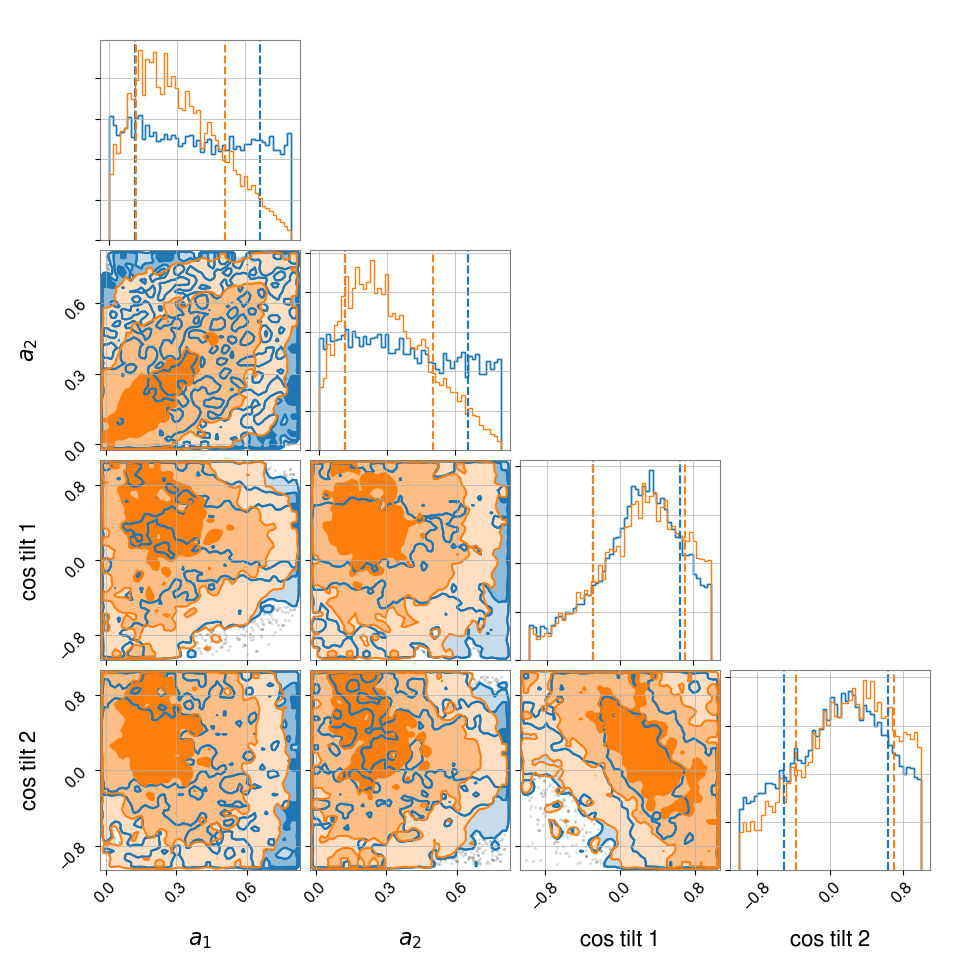}
    \end{subfigure}
    \caption{Posterior distributions for GW170814. Blue contours are calculated with uninformative priors while the orange contours are calculated using a posterior predictive distribution.}
    \label{fig:ppd_GW170814}
\end{figure*}

\begin{figure*}
    \begin{subfigure}[t]{\columnwidth}
    \includegraphics[width=0.79\linewidth]{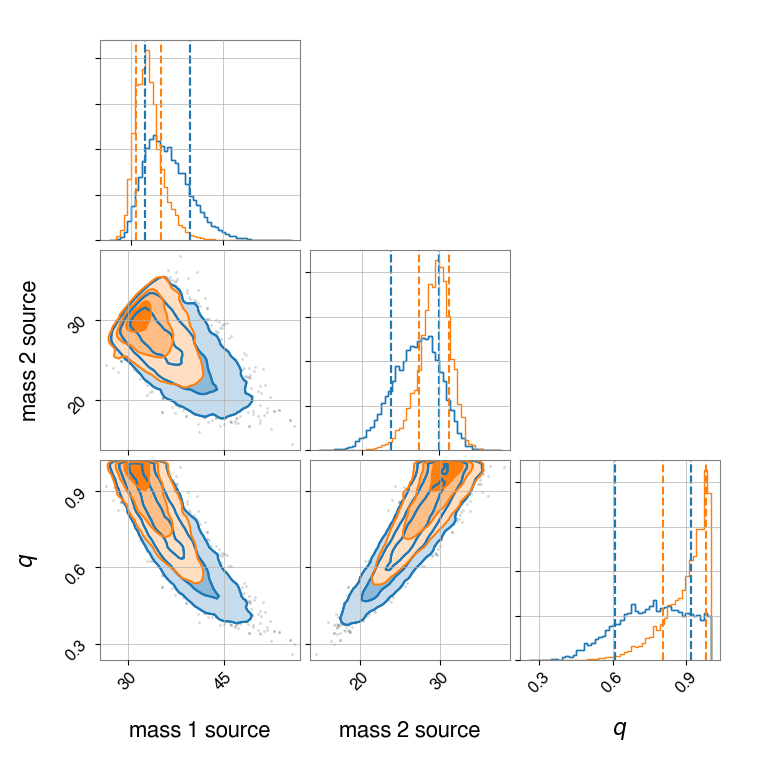} 
    \end{subfigure}
    \begin{subfigure}[t]{\columnwidth}
    \includegraphics[width=\linewidth]{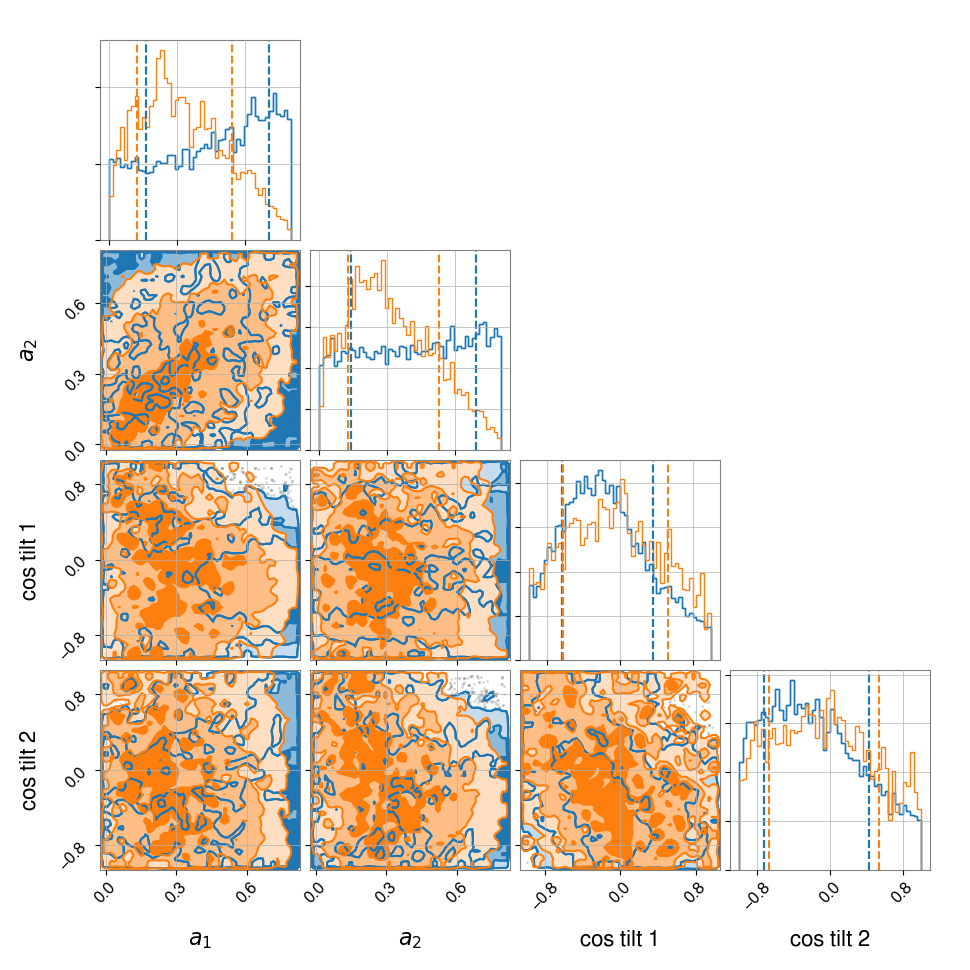}
    \end{subfigure}
    \caption{Posterior distributions for GW170818. Blue contours are calculated with uninformative priors while the orange contours are calculated using a posterior predictive distribution.}
    \label{fig:ppd_GW170818}
\end{figure*}

\begin{figure*}
    \begin{subfigure}[t]{\columnwidth}
    \includegraphics[width=0.8\linewidth]{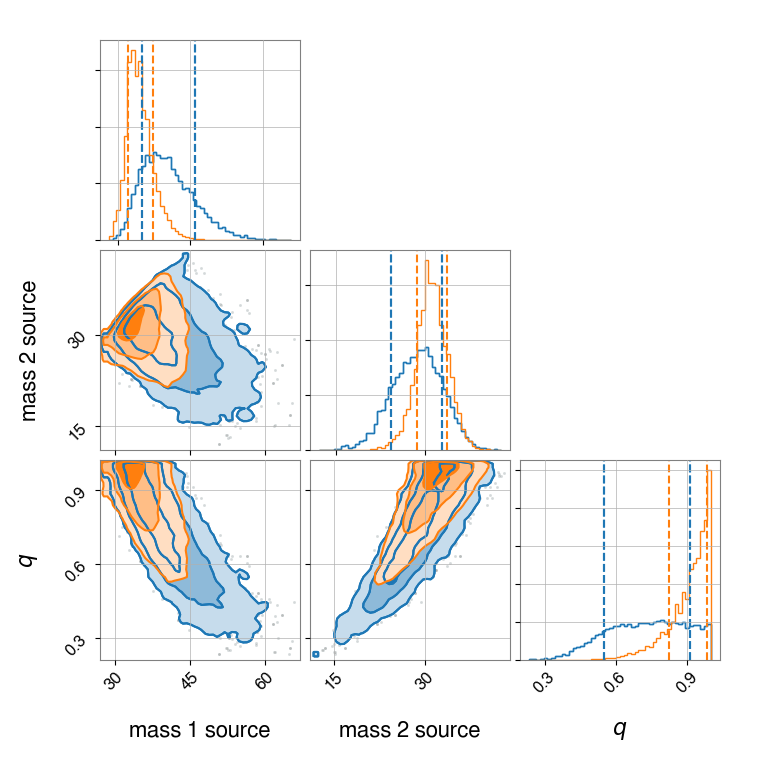} 
    \end{subfigure}
    \begin{subfigure}[t]{\columnwidth}
    \includegraphics[width=\linewidth]{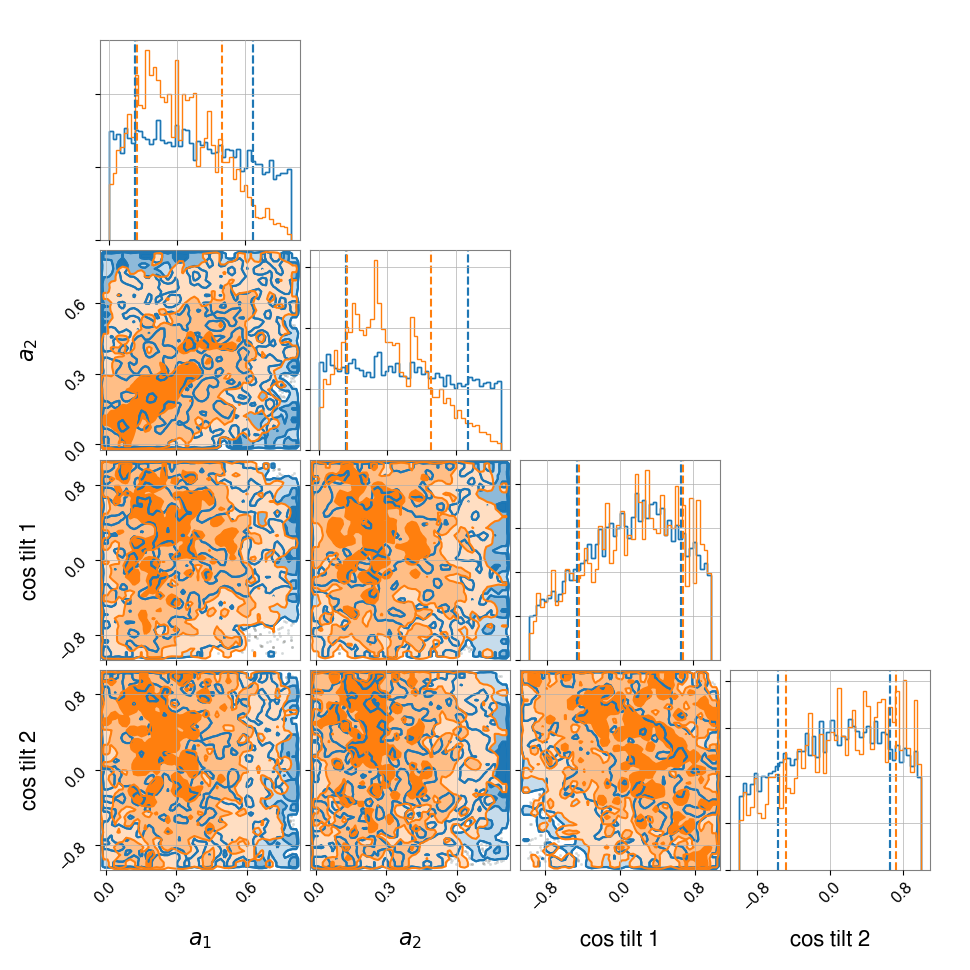}
    \end{subfigure}
    \caption{Posterior distributions for GW170823. Blue contours are calculated with uninformative priors while the orange contours are calculated using a posterior predictive distribution.}
    \label{fig:ppd_GW170823}.
\end{figure*}

In this Figs.~\ref{fig:ppd_GW150914}-\ref{fig:ppd_GW170823} we present the posteriors for the events in the GWTC-1 catalog \cite{gwtc1} calculated using an astrophysically-motivated prior distribution (following the prescription in~\ref{leave-one-out}).
The event GW170729 is included in the main body of the text in Fig.~\ref{fig:ppd_GW170729}.
The blue contours are calculated with uninformative priors while the orange contours are calculated using a posterior predictive distribution. We note that the population-informed posteriors systematically shift towards equal mass, $q=1$.
The deviations in the spin parameters are mostly driven by the prior used for the population parameters.

\end{appendix}

\end{document}